\documentclass[12pt]{iopart}
\usepackage{iopams}

\usepackage{graphicx}
\usepackage{color}
\usepackage{pict2e}

\newcommand{\bea}{\begin{eqnarray}}
\newcommand{\eea}{\end{eqnarray}}
\newcommand{\beq}{\begin{equation}}
\newcommand{\eeq}{\end{equation}}

\newcommand{\eqref}[1]{(\ref{#1})}

\newcommand{\Kerr}{{\mbox{\tiny Kerr}}}
\newcommand{\IZ}{{\mbox{\tiny IZ}}}
\newcommand{\NZ}{{\mbox{\tiny NZ}}}
\newcommand{\FZ}{{\mbox{\tiny FZ}}}

\begin{document}

\title[Asymptotically Matched Spacetime Metric]
{Asymptotically Matched Spacetime Metric for Non-Precessing, Spinning Black Hole Binaries}

\author{Louis Gallouin$^{1,2}$,
Hiroyuki Nakano$^2$,
Nicol\'{a}s Yunes$^3$
and
Manuela Campanelli$^2$}

\address{$^1$ICFP, D\'epartement de Physique de l'ENS , 24 rue Lhomond, 75005 Paris, France}

\address{$^2$Center for Computational Relativity and Gravitation,
and School of Mathematical Sciences, Rochester Institute of
Technology, 85 Lomb Memorial Drive, Rochester, New York 14623, USA}

\address{$^3$Department of Physics, Montana State University,
Bozeman, Montana 59717, USA}

\begin{abstract}

We construct a closed-form, fully analytical $4$-metric that approximately
represents the spacetime evolution of non-precessing, spinning black hole
binaries from infinite separations up to a few orbits prior to merger. We
employ the technique of asymptotic matching to join a perturbed Kerr
metric in the neighborhood of each spinning black hole to a near-zone, post-Newtonian
metric farther out. The latter is already naturally matched to a far-zone,
post-Minkowskian metric that accounts for full temporal retardation.
The result is a $4$-metric that is approximately valid everywhere in space and 
in a small bundle of spatial hypersurfaces. We here restrict our attention to 
quasi-circular orbits, but the method is valid for any orbital motion or physical scenario, 
provided an overlapping region of validity or buffer zone exists.
A simple extension of such a metric will allow for future studies of the accretion 
disk and jet dynamics around spinning back hole binaries.

\end{abstract}

\maketitle

\section{Introduction}

A closed-form, analytic understanding of the two-body problem in General
Relativity has proven quite elusive. From the point of view of gravity,
this holds true for Newtonian stars described by Newtonian gravity, but
even more so for binary compact objects, such as black holes (BHs) and 
neutron stars (NSs), in the last stages of inspiral and merger. 
Perhaps, this can be traced to the fact that
the ``two-body problem'' in General Relativity is really a ``one-spacetime''
problem, where the orbital dynamics cannot be easily separated from the
spacetime dynamics. Fortunately, the breakthroughs in numerical
relativity~\cite{Pretorius:2005gq, Campanelli:2005dd,Baker:2005vv}
have now made possible to fully compute the dynamics of any binary BH
(BBH) spacetime for a wide variety of mass ratios and spins parameters.
However, these calculations are still very expensive, and therefore are limited
to few orbital evolutions with the BHs starting at relatively close separations.

A closed-form, analytic understanding of the full spacetime in
the inspiral regime would be extremely useful to study a variety
of astrophysical phenomena that require a large number of orbits.
The study of the dynamics of accretion disks around BBHs and
of any associated electromagnetic emission and jets is a good example. 
With this goal in mind, we develop an asymptotically matched
spacetime metric for non-precessing, spinning BBHs.
Further motivation for our work is described in~\cite{Noble:2012xz}.

Our work builds on important previous analytic perturbative techniques, which have
been successfully developed to tackle the BBH problem. Post-Newtonian (PN) theory
(see, e.g.,~\cite{Blanchet:2002av} for a review) can be successfully used to describe
the motion of the BBH in the early stages of the inspiral. Here, all fields are perturbatively
expanded in a slow motion $v/c \ll 1$ and weak field $GM/(r c^2) \ll 1$
approximation~\footnote{Here, $c$ and $G$ are the speed of light and Newton's
gravitational constant, while $v$, $M$ and $r$ are the characteristic velocity,
mass and size or separation of the system. Henceforth, we will adopt the geometric
unit system, where $G=c=1$, with the useful conversion factor $1 M_{\odot} =
1.477 \; {\rm{km}} = 4.926 \times 10^{-6} \; {\rm{s}}$}. Close to each of the
BHs, one can use perturbation theory, where all fields are treated as small
deformations of a known analytic solution, such as the Schwarzschild or Kerr metric.

\begin{figure}[!ht]
\begin{center}
\includegraphics[width=0.48\textwidth,clip=true]{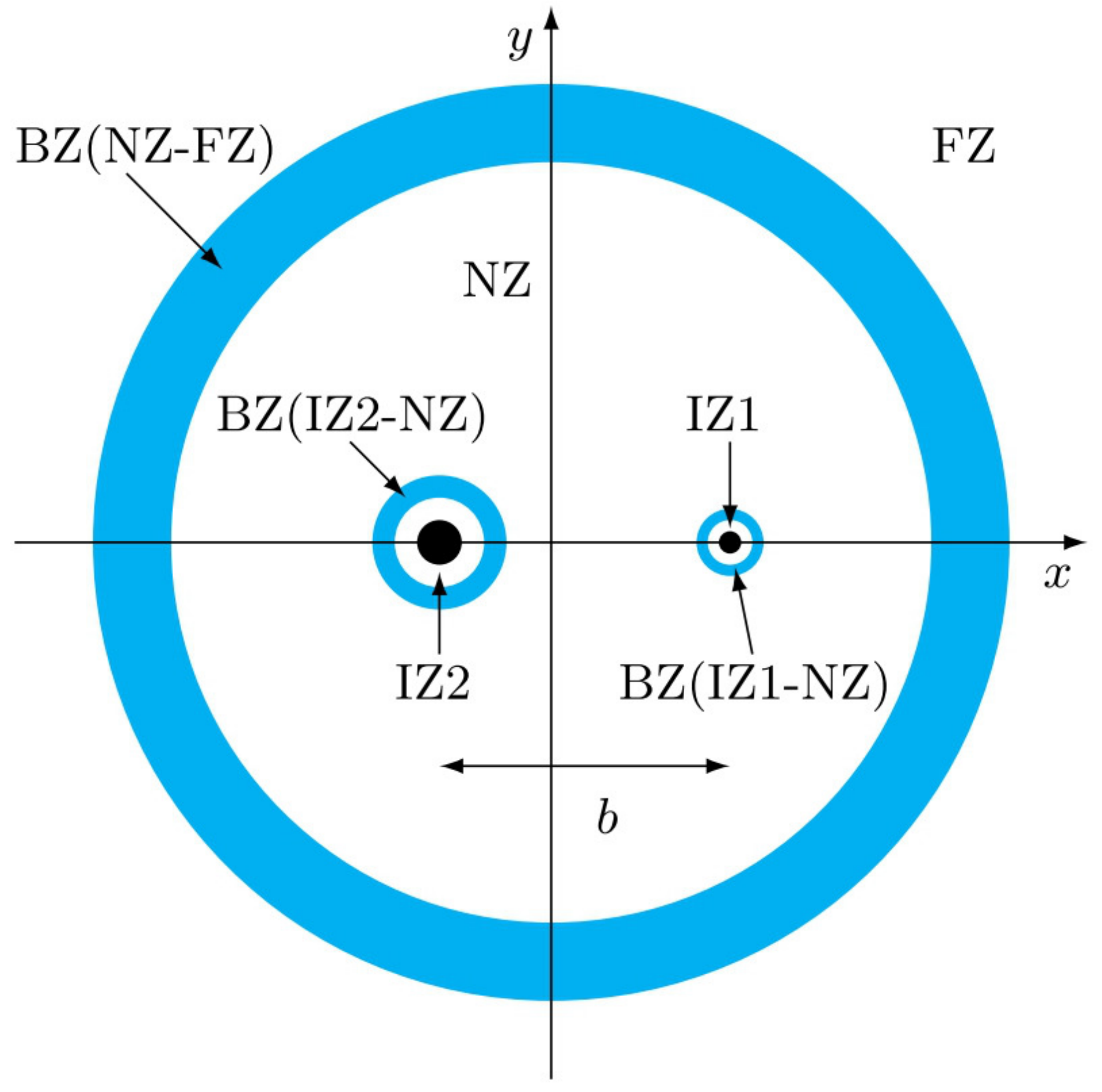}
\end{center}
\caption{BH1 and BH2 are denoted by solid black dots, where the
orbital separation is $b$. The BZs are denoted with cyan shells, the outer one representing
the FZ/NZ BZ and the two inner ones representing the NZ/IZ BZs
(see also Table~\ref{tab:zones}). The circular nature of these buffers zones is
because the diagram is schematic; in practice, one expects these to be distorted.
The IZ, NZ and FZ are also shown in the figure.}
\label{fig:zones}
\end{figure}

During the late stages of a binary BH, however,
the whole spacetime cannot be described using a single perturbative method.
For clarity, let us classify different spatial regions on a spacelike hypersurface
into zones~\cite{Thorne:1980ru} (see Figure~\ref{fig:zones} and Table~\ref{tab:zones} for a BBH).
The {\emph{inner zone}} (IZ) is defined as the region close enough to either BH
that the metric can be treated as a perturbation of the Kerr spacetime due to some external
universe~\cite{Poisson:2005pi,Taylor:2008xy,Comeau:2009bz,Poisson:2009qj,Vega:2011ue,Poisson:2011nh}.
The {\emph{near zone}} (NZ) is defined as the region both far enough from
either BH that the metric can be perturbatively expanded in a weak-field
approximation, while simultaneously much smaller than a gravitational wavelength,
so that time retardation can be treated perturbatively~\cite{Blanchet:2002av}.
The {\emph{far zone}} (FZ) is defined as the region sufficiently
far away from the center of mass of the system (much farther than a gravitational wavelength)
that the metric can be modeled via a multipolar post-Minkowskian
expansion~\cite{Blanchet:2002av}. In each of these zones, one can construct
an approximate $4$-metric, in some coordinate system well-adapted to that particular zone,
and in terms of certain physical parameters, like the BH masses and spins.

Asymptotic matching is the mathematical technique that allows one to relate the
coordinates and parameters used in adjacent zones inside of a common
overlapping region of validity, or {\emph{buffer zone}} (BZ) (see
Figure~\ref{fig:zones} and Table~\ref{tab:zones}). Formally, asymptotic
matching requires that one set the asymptotic expansion of approximate
$4$-metrics inside the BZ equal to each other. This yields a system of
differential and algebraic equations for the coordinate and parameter
transformation that relates adjacent approximate metrics. After applying such a
transformation, one can then join the approximate metrics with carefully
constructed transition functions to obtain an approximate global metric.

Such an approximate global metric is ideal as initial data, because in practice
asymptotic matching is carried in the neighborhood of some fiducial time
$t_{0}$, i.e., the BZ is the product of a $3$-torus with a small segment of the
real (temporal) line. Alvi~\cite{Alvi:1999cw,Alvi:2003pn} was the first to attempt such
an initial data construction, but ended up carrying out {\emph{asymptotic
patching}} rather than matching~\footnote{When patching, one sets the metrics
equal to each other at a point, instead of in an entire BZ region (for more
details see \cite{Yunes:2005nn,Bender}).}.
Yunes et al.~\cite{Yunes:2005nn,Yunes:2006iw,Yunes:2006mx,JohnsonMcDaniel:2009dq}
succeeded in carrying out matching for non-spinning BBHs and this data has
recently been evolved in~\cite{Reifenberger:2012yg}
(see also \cite{Chu:thesis} for numerical evolutions of superposed tidally
perturbed BHs).

\begin{table}[!ht]
  \caption{$r_{\rm in}$ and $r_{\rm out}$
  denote the approximate inner and outer boundary radius, respectively,
  while ``Region'' stands for the spatial domain in which the BZs exist.
  $r_A$ is the distance from a $A$th BH with mass $m_A$,
  $r$ is the distance from the center of mass, and
  $b$ and $\lambda$ are the orbital separation
  and the gravitational wavelength, respectively.}
  \label{tab:zones}
\begin{center}
  \begin{tabular}{l|ccc}
\br
  Zone           & $r_{\rm in}$ & $r_{\rm out}$ & Region \\
  \hline
  IZ BH1 ($r_1$)    & $0$          & $\ll b$  & $\cdot$\\
  IZ BH2 ($r_2$)    & $0$          & $\ll b$  & $\cdot$\\
  NZ ($r_A$)         & $\gg m_A$    & $\ll \lambda$ & $\cdot$\\
  FZ ($r$)            & $\gg b$ & $\infty$ & $\cdot$ \\
  IZ-NZ BZ & $\cdot$ & $\cdot$ & $m_A \ll r_A \ll b$ \\
  NZ-FZ BZ & $\cdot$ & $\cdot$ & $b \ll r \ll \lambda$\\
\br
  \end{tabular}
\end{center}
\end{table}

In this paper, we extend the calculation in~\cite{Yunes:2005nn,Yunes:2006iw,JohnsonMcDaniel:2009dq}
to non-precessing, spinning BBHs in a quasi-circular orbit, with BH spins aligned
or anti-aligned with the orbital angular momentum. The NZ and FZ are modeled
via the standard PN
approximation~\cite{Owen:1997ku,Blanchet:1998vx,Tagoshi:2000zg,Faye:2006gx,Blanchet:2006gy}
(and references therein).
The IZ around each BH is modeled as a Kerr spacetime with a leading-order vacuum
perturbation, following Yunes and Gonz\'alez~\cite{Yunes:2005ve}. The IZ metrics
are cast in a coordinate system that is both horizon-penetrating and
harmonic~\cite{Cook:1997qc}. This simplifies the matching and allows for
excision of the singularities.

We first carry out asymptotic matching in the non-spinning limit. Our results
for the matching coordinate and parameter transformation differ from those
of~\cite{Yunes:2005nn,Yunes:2006iw,JohnsonMcDaniel:2009dq} because our IZ
metrics differ by a gauge transformation. We then carry out asymptotic matching
for an aligned or counter-aligned, arbitrarily spinning binary. The matching coordinate
and parameter transformation are not affected by spin to the matching order studied.
Asymptotically matching the metrics to higher order would introduce spin-dependent
terms in the coordinate and parameter transformation, but this would require perturbed Kerr
metrics valid to higher multipolar order in the IZs.  Once asymptotic matching has been carried out,
we stitch the different metrics via appropriate transition functions,
thus obtaining a global approximate metric.

We verified that this metric is indeed an approximate solution to the Einstein
equations. One might be worried that the stitching procedure introduced errors
in the global metric that are larger than those inherently contained in any of the
approximate metrics. This is not the case because the transition functions used
satisfy the Frankenstein theorems of~\cite{Yunes:2006mx}.
We verified this qualitatively by visually inspecting different components
and the volume element of the global metric. We then
verified this quantitatively by evaluating the Ricci scalar for the global
metric on a $t = 0$ spatial hypersurface. We find that the global metric is an
approximate solution to the Einstein equations for all values of spin
considered, provided the binary orbital separation is sufficiently large,
so that BZs in which to carry out asymptotic matching exist.
We further verified that as this orbital separation is increased,
the satisfaction of the Einstein equation increases at the expected rate
(given by the matching order).

Such a global metric is technically valid close to a $t=0$ spatial hypersurface
because the BZ in which matching is carried out is formally a small $4$-volume
about this hypersurface. One can, however, extend this global metric to a
dynamical global representation of the full spacetime, following the procedure
in~\cite{Bruno_prep}. This scheme essentially consists of using a sequence of
temporally-spaced global metrics and properly gluing these together.
Such a spacetime was successfully employed to model how the inspiral
of BBHs affects accretion disks~\cite{Noble:2012xz}.

The remainder of this paper is organized as follows.
In~\sref{sec:prep}, we summarize how the metrics of each zone are
constructed. In~\sref{sec:match}, we first redo asymptotic matching
between IZ and NZ metrics in the non-spinning case but using an IZ metric
different from the one used
in~\cite{Yunes:2005nn,Yunes:2006iw,Yunes:2006mx,JohnsonMcDaniel:2009dq},
and then extend it to the spinning case.
We present some technical details and supplementary analyses in the appendices.
In~\ref{app:comp}, we review how this paper is related to~\cite{Yunes:2005ve},
and extend the formulation by including the $m=0$ azimuthal mode.
In~\ref{app:fullV}, we present a detailed and complete analysis of
asymptotic matching. In~\ref{app:trans}, we discuss how to glue the
metrics in each zone with transition functions. In~\ref{app:higher},
we summarize some higher order metrics which are available 
for the non-spinning case in the supplemental material of~\cite{JohnsonMcDaniel:2009dq}.

All throughout, we use the following conventions, following mostly Misner,
Thorne and Wheeler~\cite{MTW}. We use the Greek letters $(\alpha, \beta,
\cdots)$ to denote spacetime indices, and Latin letters $(i, j, \cdots)$
to denote spatial indices. The metric is denoted $g_{\mu \nu}$ and
it has signature $(-,+,+,+)$. We use geometric units, with $G=c=1$.

\section{Approximate Metrics}\label{sec:prep}

\subsection{Inner Zone}\label{subsec:IZ}

The metric in either IZ is approximated with a Kerr solution plus a linear
vacuum perturbation~\cite{Yunes:2005ve}:
\begin{eqnarray}
g_{\mu \nu}^{\IZ} = g_{\mu \nu}^{\Kerr} + h_{\mu \nu}^{\IZ}\,;
\end{eqnarray}
the construction of $h_{\mu \nu}^{\IZ}$ is difficult but essential for our
purposes. Vacuum perturbations of Schwarzschild have been studied, for
example, in~\cite{Detweiler:2005kq}, where one can use the
Regge-Wheeler-Zerilli-Moncrief
formalism~\cite{Regge:1957td,Zerilli:1971wd,Moncrief:1974am}. When considering
the Kerr background, however, the perturbation equations do not easily
decouple, as tensor spherical harmonics are not eigenfunctions of the angular
sector of the linearized Einstein equations~\cite{Teukolsky:1973ha}. Instead
one has to carry out the so-called Chrzanowski
procedure~\cite{Chrzanowski:1975wv}, amended by Wald~\cite{Wald:1978vm} and by
Kegeles and Cohen~\cite{Kegeles:1979an}, as we describe below.

In the Chrzanowski procedure~\cite{Chrzanowski:1975wv}, the vacuum perturbation
$h_{\mu\nu}^{\IZ}$ is obtained from a so-called Hertz potential $\Psi$ via
\begin{eqnarray}
h_{\mu\nu}^{\IZ} &= \hat{h}_{\mu\nu} \left[\Psi\right] \,,
\end{eqnarray}
where $\hat{h}_{\mu\nu}\left[\cdot\right]$ is a differential
operator. This potential must satisfy a certain non-linear differential
equation with a source given by the Newman-Penrose scalar $\psi_{0}$ (or
$\psi_{4}$ depending on the radiation gauge used). This differential relation
can be inverted through the Teukolsky-Starobinski relation to yield $\Psi$ in
terms of $\psi_{0,4}$ (\cite{Wald:1978vm,Lousto:2002em,Ori:2002uv}). Thus, the
construction of a metric perturbation reduces to finding an appropriate
solution for the Newman-Penrose scalar.

The Newman-Penrose scalar must, of course, satisfy the Teukolsky equation, but
when the source is a slowly-varying external universe (as is the case when the
perturbation is due to a BH binary companion in a circular orbit with large
semi-major axis), this equation can be solved perturbatively. One first
performs a harmonic decomposition of $\psi_{0}$ in terms of spin-weight $2$
spherical harmonics (or $\psi_{4}$ in terms of spin-weight $-2$ spherical
harmonics), $_{2}Y^{\ell m}$:
\begin{equation}
\psi_{0} = \sum_{\ell,m} R_{\ell m}(r) z_{\ell m}(v) \; {}_{2}Y^{\ell m}(\theta,\phi)\,,
\end{equation}
where the radial and temporal dependence are product decomposed in terms of
unknown real functions $R_{\ell m}(r)$ and unknown complex functions $z_{\ell m}(v)$,
where $v$ is the advanced Kerr-Schild time coordinate. The latter can be written
in terms of certain combinations of electric and magnetic tidal tensor components,
which characterize the perturbations of the external universe.
To leading order in a small-hole/slow-motion approximation~\cite{Poisson:2004cw}
in BH perturbation theory, it suffices to keep only the $\ell=2$
quadrupolar deformation (see also~\ref{app:comp}).
The radial functions are then required to satisfy the (time-independent) Teukolsky equation,
which one can solve in terms of hypergeometric functions~\cite{Poisson:2004cw,Yunes:2005ve}.
This means that $z_{\ell m}$ is constant or slowly varying in time.
With the Newman-Penrose scalars under control, one then proceeds to compute
the Hertz potential $\Psi$, and from that, the metric perturbation.
Reference~\cite{Yunes:2005ve} provides explicit expressions for
$g_{\mu \nu}^{\IZ}$ in Boyer-Lindquist (BL) coordinates.

For ease during the matching procedure, one wishes to express the IZ metric
in a coordinate system that closely parallels that used in the NZ. We thus transform
the IZ metric from BL coordinates ($t_{\rm BL},\,r_{\rm BL},\,\theta_{\rm BL},\,\phi_{\rm BL}$)
to harmonic, horizon-penetrating coordinates ($T,\,X,\,Y,\,Z$). Harmonic coordinates
are those that satisfy $\square x^{\mu} = 0$, which of course defines a class of coordinate systems.
Several coordinate transformations between BL and different members of this class were developed
in~\cite{Abe:1987ms,Ruiz:1986re,Cook:1997qc,Aguirregabiria:2001vk,Cook:2000vr,Hergt:2007ha,Sopuerta:2011te};
we here employ one such transformation that leads to harmonic coordinates
that are also horizon-penetrating~\cite{Cook:1997qc}, namely 
\begin{eqnarray}
\fl
T = t_{\rm BL} + \frac{r_{+}^2+A^2}{r_{+}-r_{-}}
\,\ln \left|\frac{r_{\rm BL}-r_{+}}{r_{\rm BL}-r_{-}}\right|
\,,
\;
Z = (r_{\rm BL} - M) \cos \theta_{\rm BL} \,,
\nonumber \\
\fl
X + i\,Y = (r_{\rm BL}-M+i\,A) e^{i\,\phi_{\rm IK}} \,\sin \theta_{\rm BL} \,;
\;
\phi_{\rm IK} = \phi_{\rm BL} + \frac{A}{r_{+}-r_{-}}
\,\ln \left|\frac{r_{\rm BL}-r_{+}}{r_{\rm BL}-r_{-}}\right|
\,,
\label{eq:longCT}
\end{eqnarray}
where $A = S/M$ is the dimensional Kerr spin parameter associated with the Kerr background with spin $S$,
and $M$ is the mass of the Kerr background, with $r_{\pm} = M \pm \sqrt{M^2-A^2}$
the location of the unperturbed inner and outer horizons in BL coordinates.
The quantity $\phi_{\rm IK}$ should be thought of as an azimuthal variable of ingoing Kerr (IK) coordinates,
not as a spherical polar $\Phi$ coordinate associated with harmonic,
horizon-penetrating coordinates.

The IZ metric is then in harmonic, horizon-penetrating coordinates $X^{\alpha}$
and it is characterized by the parameters $\Lambda^{\alpha} = (M,A,z_{R,m},z_{I,m})$,
where the first two are associated with the background and the latter two are related
to the metric perturbation, i.e., the real and imaginary parts of $z_{2 m}$.
In~\sref{sec:match}, we will carry out asymptotic matching and relate
these coordinates and parameters to the NZ ones.

\subsection{Near Zone}

The NZ metric is chosen to be given by the PN expansion,
\begin{equation}
g_{\mu \nu}^{\NZ} = \eta_{\mu \nu} + h_{\mu \nu}^{\NZ}\,,
\end{equation}
where $\eta_{\mu \nu}$ is the Minkowski metric, while $h_{\mu \nu}$ is a PN metric perturbation.
The latter can be decomposed into spin-independent and spin-dependent terms. All spin-independent
terms are given explicitly in~\cite{Blanchet:1998vx},
and we will here only keep the first non-vanishing spin terms
in the metric perturbation, which can be found in~\cite{Tagoshi:2000zg,Faye:2006gx}.

The metric perturbation of linear-in-spin contributions can be written as
\begin{eqnarray}
\fl
\delta h_{00}^{\NZ, \rm (S)} = 2 \delta V^{\rm (S)} + {\cal{O}}(v^7) \,,
\quad
\delta h_{0i}^{\NZ, \rm (S)} = -4 \delta V_i^{\rm (S)} + {\cal{O}}(v^6) \,,
\quad
\delta h_{ij}^{\NZ, \rm (S)} = {\cal{O}}(v^5) \,.
\label{eq:NZspinC}
\end{eqnarray}
The PN ordering system we employ is as follows:
a term that corrects the leading-order coefficient of some expression by a quantity
of relative  ${\cal{O}}(v^{2n}) = {\cal{O}}(v^{2n}/c^{2n})$ is said to be of $n$-th PN order.
The spin potentials to leading order are
\begin{eqnarray}
\delta V^{\rm (S)} = \frac{2}{r_1^2} \epsilon_{ijk} v_1^i s_1^j n_1^k
+ (1 \leftrightarrow 2) + {\cal{O}}(v^5) \,,
\nonumber \\
\delta V_i^{\rm (S)} = \frac{1}{2 r_1^2} \epsilon_{ijk} s_1^j n_1^k
+ (1 \leftrightarrow 2) + {\cal{O}}(v^3) \,.
\end{eqnarray}
Here, $s_A^i$ denotes the spin angular momentum of the $A$th PN particle,
which has dimensions of (mass)$^2$, while the $A$th particle's mass,
location and velocity are given by $m_{A}$, $y_{A}^{i}$ and $v_{A}^{i}$ respectively.
We also use the notation, $r_A=|\mathbf{x}-\mathbf{y}_A|$ and $n_A^i=(x^i - y_A^i)/r_A$.

Putting all of this together, one has the 1.5PN order NZ metric,
\begin{eqnarray}
\fl
g_{00}^{\NZ} + 1 =&
\frac{2 m_{1}}{r_{1}} + \frac{m_{1}}{r_{1}} \left[4 \mathbf{v}_{1}^{2} - (\mathbf{n}_{1} \cdot \mathbf{v}_{1})^{2} \right]
- 2 \frac{m_{1}^{2}}{r_{1}^{2}} - m_{1} m_{2} \Biggl[ \frac{2}{r_{1} r_{2}} + \frac{r_{1}}{2 b^{3}}
- \frac{r_{1}^{2}}{2 r_{2} b^{3}} + \frac{5}{2 r_{2} b} \Biggr]
\nonumber \\
\fl &
+ \frac{4 m_{1} m_{2}}{3 b^{2}} (\mathbf{n}_{12} \cdot \mathbf{v}_{12})
+ \frac{4}{r_1^2} \epsilon_{ijk} v_1^i s_1^j n_1^k + (1 \leftrightarrow 2) + {\cal{O}}(v^6)
\,,
\nonumber \\
\fl
g_{0i}^{\NZ} =
& \hspace{-10mm}
-\frac{4 m_{1}}{r_{1}} v_{1}^{i} 
-\frac{2}{r_1^2} \epsilon_{ijk} s_1^j n_1^k
+ (1 \leftrightarrow 2) + {\cal{O}}(v^5)
\,,
\nonumber \\
\fl
g_{ij}^{\NZ} - \delta_{ij} =& \frac{2 m_{1}}{r_{1}} \delta_{ij}
+ (1 \leftrightarrow 2) + {\cal{O}}(v^4)
\,,
\label{eq:NZmetric}
\end{eqnarray}
where we have introduced the notation
\begin{eqnarray}
b = |\mathbf{y}_{1}-\mathbf{y}_2| \,,
\quad
\mathbf{n}_{12} = (\mathbf{y}_{1} - \mathbf{y}_{2})/b \,,
\quad
\mathbf{v}_{12} = \mathbf{v}_{1} - \mathbf{v}_{2}
\,.
\end{eqnarray}
The quantity $b$ is the same as the commonly used $r_{12}$ in the PN literature.
Quadratic spin term are here neglected, as they enter at higher PN order.
The spin-independent terms are available up to 2.5PN order from~\cite{Blanchet:1998vx},
and implemented as supplemental material in~\cite{JohnsonMcDaniel:2009dq}.
We briefly summarize these higher order terms in~\ref{app:higher}.

For quasi-circular orbits, certain simplifications are possible. First, we have
$\mathbf{n}_{12} \cdot \mathbf{v}_{1,2,12} = {\cal{O}}(v^5)$, since the orbit radially decays
due to gravitational radiation reaction. Ignoring the latter,
$\mathbf{b} = b \,\{\cos\omega t,\,\sin\omega t,\,0 \}$, where
\begin{eqnarray}
\omega &= \sqrt\frac{m}{b^{3}} \left[ 1 + \frac{m}{2b}\left(\frac{m_{1} m_{2}}{m^{2}} - 3\right)
+ {\cal{O}}(v^4) \right] \,,
\end{eqnarray}
is the orbital angular frequency. Here, $m= m_{1} + m_{2}$ is the total mass of the system.
In \sref{sec:match}, we will use the positions of the holes,
\begin{eqnarray}
\mathbf{y}_{1} = \frac{m_{2}}{m} \mathbf{b} + {\cal{O}}(v^{4}) \,,
\quad
\mathbf{y}_{2} = - \frac{m_{1}}{m} \mathbf{b} + {\cal{O}}(v^{4}) \,,
\end{eqnarray}
that are obtained in the calculation of the center of mass of the system.
Higher-order expressions for these quantities are given later in Eq.~(\ref{eq:relCOM}).

The NZ metric is then given in harmonic PN coordinates $x^{\alpha}$,
and characterized by the parameters $\lambda^{\alpha} = (m_{1},m_{2}, b, s_{1}^{i}, s_{2}^{i})$.
In~\sref{sec:match}, we will carry out asymptotic matching
and relate these coordinates and parameters to the IZ ones.

\subsection{Far Zone}

The FZ metric can be obtained via the direct integration of the relaxed Einstein
equations~\cite{Will:1996zj,Pati:2000vt,Pati:2002ux} in harmonic gauge,
or alternatively via the multipolar formalism of Blanchet, Damour, and Iyer
(BDI)~\cite{Blanchet:1995fg,Blanchet:1995fr,Blanchet:1996wx,Blanchet:2002av}.
The FZ metric in~\cite{Will:1996zj,Pati:2002ux,JohnsonMcDaniel:2009dq} is given by
\begin{eqnarray}
g_{00}^{\FZ} = -\left[1 - \frac{1}{2}\, h^{00}_{\FZ} + \frac{3}{8}\, \left(h_{\FZ}^{00}\right)^{2}  \right]
+ \frac{1}{2}\, h_{\FZ}^{kk} + {\cal{O}}(v^6) \,,
\nonumber \\
g_{0k}^{\FZ} = - h_{\FZ}^{0k} + {\cal{O}}(v^5) \,,
\quad
g_{kl}^{\FZ} = \left[1 + \frac{1}{2}\, h_{\FZ}^{00} \right]\delta^{kl} + {\cal{O}}(v^4) \,,
\label{eq:FZ1}
\end{eqnarray}
where the metric potentials $h_{\FZ}^{\mu \nu}$ are 
\begin{eqnarray}
\fl
h_{\FZ}^{00} =
4\, \frac{{\cal{I}}}{r}
+ 2\, \partial_{kl} \left[ \frac{{\cal{I}}^{kl}(u)}{r} \right]
- \frac{2}{3}\, \partial_{klm} \left[ \frac{{\cal{I}}^{klm}(u)}{r} \right]
+ 7\, \frac{{\cal{I}}^{2}}{r^{2}} + {\cal{O}}(v^{6}) \,,
\nonumber \\
\fl
h_{\FZ}^{0k} =
- 2\, \partial_{l} \left[ \frac{\dot{\cal{I}}^{kl}(u)}{r} \right]
+ 2\, \epsilon^{lkp} \frac{n^{l} {\cal{J}}^{p}}{r^{2}} + {\cal{O}}(v^{5}) \,,
\nonumber \\
\fl
h_{\FZ}^{kl} = 2\,  \frac{\ddot{\cal{I}}^{kl}(u)}{r}
- \frac{2}{3}\, \partial_{p} \left[ \frac{\ddot{\cal{I}}^{klp}(u)}{r} \right]
- \frac{8}{3}\, \epsilon^{ps(k|} \partial_{s} \left[ \frac{\dot{\cal{J}}^{p|l)}(u)}{r} \right]
+ \frac{{\cal{I}}^{2}}{r^{2}} \hat{n}^k\hat{n}^l + {\cal{O}}(v^{6}) \,,
\label{eq:FZ2}
\end{eqnarray}
with $r=|\mathbf{x}|$ the distance from the binary's center-of-mass
to the field point, $n^{k} := x^{k}/r$, and $u=t-r$ is the retarded time.
We follow here the PN order counting explained in~\cite{JohnsonMcDaniel:2009dq},
and we keep terms up to the same order as those retained in the NZ metric.
Higher-order terms are of course available,
and implemented for the non-spinning case as the supplemental material
in~\cite{JohnsonMcDaniel:2009dq} (see \ref{app:higher} for a brief summary).

The FZ metric is then given in terms of source multipole moments, which are corrected by spin terms.
The spin contributions to the multipole moments can be read from Eqs.~(B5) and (C1)
in~\cite{Will:2005sn} (see also \cite{Blanchet:2006gy}):
\begin{eqnarray}
\delta \mathcal{J}^k_{\rm (S)} = s_1^k + s_2^k \,,
\quad
\delta \mathcal{J}^{kl}_{\rm (S)} = \frac{1}{2}
\left( 3\, s_1^k y_1^l - \delta^{kl} s_1^m y_1^m \right)
+ (1 \leftrightarrow 2)\,,
\label{eq:FZspinC}
\end{eqnarray}
where $y_A^i$, $v_A^i$ and $s_A^i$ are the same as in the NZ but evaluated at retarded time $u$.
All non-spinning terms are given explicitly in~\cite{JohnsonMcDaniel:2009dq}.

We have verified that the NZ and FZ metrics are automatically asymptotically matched
in the BZ constructed from the intersection of the NZ and FZ. In practice, the leading 1.5PN
order spin effects in the NZ metric,
$g_{00}^{\NZ}$ and $g_{0i}^{\NZ}$ are matched to the spinning parts of $(1/2)h_{\FZ}^{kk}$
and $-h_{\FZ}^{0k}$ in the FZ metric $g_{00}^{\FZ}$ and $g_{0i}^{\FZ}$, respectively.
That is, if one asymptotically expands both of these metrics in this BZ,
then they are automatically equal to each other without requiring any coordinate or parameter
transformation.

To establish this result, one need to only worry about the following
transformation of the relative to center-of-mass coordinates.
According to Eq.~(5.5) of~\cite{Faye:2006gx},
the relative position $b \; \mathbf{n}_{12}=\mathbf{y}_1-\mathbf{y}_2$
and velocity $\mathbf{v}_{12}=\mathbf{v}_1-\mathbf{v}_2$ are converted from
the quantities in the center of mass system via
\begin{eqnarray}
\mathbf{y}_1 =& \left[\frac{m_2}{m}
+ \frac{\eta}{2}\delta \left(v^2-\frac{m}{b}\right)\right]\,b \,\mathbf{n}_{12}
+ \frac{\eta}{m}\,\mathbf{v}_{12} \times \mathbf{\Sigma}
\,,
\nonumber \\
\mathbf{y}_2 =& \left[-\frac{m_1}{m}
+ \frac{\eta}{2}\delta \left(v^2-\frac{m}{b}\right)\right]\,b \,\mathbf{n}_{12}
+ \frac{\eta}{m}\,\mathbf{v}_{12} \times \mathbf{\Sigma}
\,,
\label{eq:relCOM}
\end{eqnarray}
up to 1.5PN order. Here, $\delta=(m_1-m_2)/m$, $\eta=m_1 m_2/m^2$
and $\mathbf{\Sigma} = m ( \mathbf{S}_2/m_2 - \mathbf{S}_1/m_1 )$.

The above transformation is affected by the choice of {\emph{spin supplementary condition}} (SSC)
(see, e.g., Appendix A of~\cite{Kidder:1995zr,Will:2005sn}).
Although the NZ and FZ metrics are technically computed using different SSCs,
this is not a problem here due to the PN order to which we work.

\section{Asymptotic Matching}\label{sec:match}

In this section, we carry out asymptotic matching between IZ and NZ metrics.
We will here follow the same procedure as that first introduced in~\cite{Yunes:2005nn},
and recently refined in~\cite{JohnsonMcDaniel:2009dq}.
We first concentrate on matching the IZ metric around BH1 and the NZ metric;
matching between the other IZ metric and the NZ can be obtained later via a symmetry
transformation. In asymptotic matching, one expands both metrics in the BZ, $m_{1} \ll r_1 \ll b$
and $t \ll b$, and then requires that they be diffeomorphic to each other.
This leads to a set of differential equations that relate the coordinates used in each metric,
as well as a set of algebraic equations that relate the parameters used in each zone.
The IZ metric around BH1 depends on $5$ complex parameters
$z_{2m}$ ($m=(-2,-1,0,1,2)$), which must be determined with the matching procedure.

Before proceeding with the matching, let us briefly summarize the coordinates,
parameters and expansions which we employ on the IZ and NZ metrics.
The NZ metric is expressed in NZ, harmonic, PN coordinates $x^{\alpha}$
and depends on parameters $\lambda^{\alpha} = (m_{1},m_{2},b,s_{1}^{i},s_{2}^{i})$,
where we recall that $m_{1,2}$ are the masses of the PN particles,
$s_{1,2}^{i}$ is their spin angular momentum, and $b$ is their separation.
The IZ metric is expressed in IZ, harmonic, horizon penetrating coordinates $X^{\alpha}$
and depends on parameters $\Lambda^{\alpha} = (M,S,z_{R,m},z_{I,m})$,
where $M$ and $S$ are the mass and the magnitude of the spin angular momentum
associated with the Kerr background, while $z_{R,m}$ and $z_{I,m}$ are the real and imaginary parts
of the $10$ tidal field parameters that characterize the metric perturbation.
We expand these metrics in the BZ in powers of $(m_{2}/b)^{1/2}={\cal{O}}(v)$:
\begin{eqnarray}
X^{\alpha}\left( x^{\beta} \right) =& \sum_{i=0}^{n} \left( \frac{m_{2}}{b} \right)^{i/2}
 \left( X^{\alpha} \right)_{i} \left( x^{\beta} \right)
+ {\cal{O}}( v^{n+1} ) \,,
\nonumber \\
\Lambda^{\alpha}\left( \lambda^{\beta} \right) =& \sum_{i=0}^{n} \left( \frac{m_{2}}{b} \right)^{i/2}
 \left( \Lambda^{\alpha} \right)_{i} \left( \lambda^{\beta} \right)
+ {\cal{O}}( v^{n+1} ) \,,
\end{eqnarray}
or more explicitly
\begin{eqnarray}
M(\lambda^{\alpha}) =& \sum_{i=0}^{n} \left( \frac{m_{2}}{b} \right)^{i/2}
\left( M \right)_{i}\left(\lambda^{\beta}\right) + {\cal{O}}( v^{n+1} ) \,,
\\
S(\lambda^{\alpha}) =& \sum_{i=0}^{n} \left( \frac{m_{2}}{b} \right)^{i/2}
\left( S \right)_{i}\left(\lambda^{\beta}\right) + {\cal{O}}( v^{n+1} ) \,,
\\
\label{z-exp}
z_{R/I,m}(\lambda^{\alpha}) &= \sum_{i=0}^{n} \left( \frac{m_{2}}{b} \right)^{i/2}
\left( z_{R/I,m} \right)_{i}\left(\lambda^{\beta}\right) + {\cal{O}}( v^{n+1} ) \,.
\end{eqnarray}
Here, $(X^{\alpha})_{i}$ are $i$ functions of the NZ coordinates $x^{\beta}$,
while $ \left( \Lambda^{\alpha} \right)_{i}$ are $i$ functions
of the NZ parameters $\lambda^{\beta}$. Similarly, $(M)_{i}$, $(S)_{i}$ and $(z_{R/I,m})_{i}$
are also functions of the $\lambda^{\beta}$ NZ parameters. We here take the sums up to $n = 2$,
i.e., carry out asymptotic matching to $O[(m_{2}/b)^1]$.

With this expansion in hand, we proceed to carry out asymptotic matching in the next subsections.
We begin by focusing on the non-spinning case. This is different from the work
in~\cite{Yunes:2005nn,JohnsonMcDaniel:2009dq} because we here use the IZ metric of~\cite{Yunes:2005ve}
in the non-spinning limit. We will then proceed with the matching of the spinning case.

\subsection{Expansion of the Non-Spinning IZ and NZ Metrics}

Let us first expand the NZ and IZ metrics in the BZ. Formally, we expand the NZ metric
as in~\cite{JohnsonMcDaniel:2009dq}:
\begin{eqnarray}
g_{\alpha \beta} =& (g_{\alpha \beta})_{0}
+ \sqrt{ \frac{m_{2}}{b} }(g_{\alpha \beta})_{1}
+ \left( \frac{m_{2}}{b} \right) (g_{\alpha \beta})_{2}
+ {\cal{O}}(v^3) \,,
\end{eqnarray}
without any spin contributions, where
\begin{eqnarray}
\fl
(g^{\NZ}_{\alpha \beta})_{0} = \eta_{\alpha \beta} \,,
\quad
(g^{\NZ}_{\alpha \beta})_{1} = 0 \,,
\nonumber \\
\fl
(g^{\NZ}_{\alpha \beta})_{2} = \biggl[ \frac{2m_{1}}{m_{2}} \frac{b}{(r_{1})_{0}}
+ 2 - \frac{2}{b} \left\{(\mathbf{r}_{1})_{0} \cdot (\mathbf{\hat b})_{0}\right\}
+ \frac{1}{b^{2}} \left\{3 [(\mathbf{r}_{1})_{0} \cdot (\mathbf{\hat b})_{0}]^{2} - [(r_{1})_{0}]^{2}\right\}
\biggr]
\Delta_{\alpha \beta} \,.
\label{eq:NZ_bz}
\end{eqnarray}
We have here defined a ``lowered four dimensional Kronecker delta'',
$\Delta_{\alpha \beta} = \mbox{diag}(1,1,1,1)$,
and $(\hat b^k)_{0}=\hat x^k=\{1,\,0,\,0\}$ is a unit vector, with the assumption
that each BH is initially located on the $x$-axis.
We also use the Cartesian PN coordinate basis vectors,
$\hat t^\alpha=\{1,\,0,\,0,\,0\}$, $\hat x^\alpha=\{0,\,1,\,0,\,0\}$,
$\hat y^\alpha=\{0,\,0,\,1,\,0\}$ and $\hat z^\alpha=\{0,\,0,\,0,\,1\}$ later.

Before we expand the IZ metric in the BZ, it is convenient to re-express
the parameters $z_{R,m}$ and $z_{I,m}$ in terms of
the electric ${\cal E}_{kl}$ and magnetic ${\cal B}_{kl}$ tidal tensor components:
\begin{eqnarray}
z_{R,0} &= -2\, {\cal E}_{XX}-2\, {\cal E}_{YY} \,,
\;
z_{I,0} = -2\, {\cal B}_{XX}-2\, {\cal B}_{YY} \,,
\nonumber \\
z_{R,1} &= -2\, {\cal E}_{XZ}-2\, {\cal B}_{YZ} \,,
\;
z_{R,-1} = -2\, {\cal E}_{XZ}+2\, {\cal B}_{YZ} \,,
\nonumber \\
z_{I,1} &= 2\, {\cal E}_{YZ}-2\, {\cal B}_{XZ} \,,
\;
z_{I,-1} = -2\, {\cal E}_{YZ}-2\, {\cal B}_{XZ} \,,
\nonumber \\
z_{R,2} &= -2\, ({\cal E}_{XX}-{\cal E}_{YY}) - 4\,{\cal B}_{XY} \,,
\;
z_{R,-2} = -2\, ({\cal E}_{XX}-{\cal E}_{YY}) + 4\,{\cal B}_{XY} \,,
\nonumber \\
z_{I,2} &= 4\, {\cal E}_{XY}+2\, ({\cal B}_{YY}-{\cal B}_{XX}) \,,
\;
z_{I,-2} = -4\, {\cal E}_{XY}+2\, ({\cal B}_{YY}-{\cal B}_{XX}) \,,
\label{eq:zRItoEB}
\end{eqnarray}
where we have used the traceless conditions,
\begin{eqnarray}
{\cal E}_{XX}+{\cal E}_{YY}+{\cal E}_{ZZ}=0 \,,
\quad
{\cal B}_{XX}+{\cal B}_{YY}+{\cal B}_{ZZ}=0 \,,
\end{eqnarray}
(see also~\ref{app:comp}). This will allow for a more direct comparison
of our calculation with those of~\cite{JohnsonMcDaniel:2009dq}.

Let us now expand the IZ metric in the BZ. Since the full form of the BZ expanded IZ metric
is too long and unilluminating to be included here, we only provide terms up to the second order:
\begin{eqnarray}
(g^{\IZ}_{\alpha \beta})_{0} =& \eta_{\alpha \beta}
\,, \quad
(g^{\IZ}_{\alpha \beta})_{1} = 0
\,, \nonumber \\
(g^{\IZ}_{00})_{2} =& \frac{2(M_{1})_{0}}{m_{2}} \frac{b}{(R)_{0}}
- \frac{1}{b^{2}} (\bar{\cal E}_{kl})_{0} (X^{k})_{0} (X^{l})_{0}
\,, \nonumber \\
(g^{\IZ}_{0i})_{2} =& \frac{1}{3b^{2}} \frac{(X_{i})_{0}}{(R)_{0}} (\bar{\cal E}_{kl})_{0} (X^{k})_{0} (X^{l})_{0}
+ \frac{2}{3b^{2}} (R)_{0} (\bar{\cal E}_{ik})_{0} (X^{k})_{0}
\,, \nonumber \\
(g^{\IZ}_{ij})_{2} =& \biggl( \frac{2(M)_{0}}{m_{2}} \frac{b}{(R)_{0}}
- \frac{1}{3b^{2}} (\bar{\cal E}_{kl})_{0} (X^{k})_{0} (X^{l})_{0} \biggr) \delta_{ij}
- \frac{2}{3b^{2}} (\bar{\cal E}_{ij})_{0} (R)_{0}^{2}\,.
\label{eq:IZ_bz}
\end{eqnarray}
Recalling here that the tidal tensors must also be expanded as in Eq.~\eqref{z-exp},
we have
\begin{eqnarray}
{\cal E}_{kl} &= \frac{m_{2}}{b^{3}} (\bar{\cal E}_{kl})_{0} + {\cal{O}}(v^{3})
\,,
\quad
{\cal B}_{kl} &= \left( \frac{m_{2}}{b} \right)^{3/2} \frac{1}{b^{2}} (\bar{\cal B}_{kl})_{0} + {\cal{O}}(v^{4})
\,,
\end{eqnarray}
to leading order (see also Eq.~(5.3) in~\cite{JohnsonMcDaniel:2009dq}).
Since ${\cal B}_{kl}$ is higher order than ${\cal E}_{kl}$, it can be ignored
when matching to leading order. This metric is nonsingular at the horizon $R = M$
due to the use of harmonic, horizon-penetrating coordinates;
this is also true in the spinning case. Therefore, for sufficiently small perturbations,
the horizon-penetrating character of the metric is preserved.

\subsection{Matching in the Non-spinning Case}

We carry out asymptotic matching order by order in $(m_{2}/b)^{1/2}$
and use the same notation as in~\cite{JohnsonMcDaniel:2009dq}.

\subsubsection{Zeroth-Order Matching: $O[(m_{2}/b)^0]$}

At zeroth order, we have
\begin{eqnarray}
(g^{\NZ}_{\alpha \beta})_{0} &= (A_{\alpha}{}^{\gamma})_{0} (A_{\beta}{}^{\delta})_{0} (g^{\IZ}_{\gamma \delta})_{0}
\,,
\end{eqnarray}
with $A_{\alpha}{}^{\beta} = \partial_{\alpha} X^{\beta}$.
Because $(g^{\NZ}_{\alpha \beta})_{0} = (g^{\IZ}_{\alpha \beta})_{0} = \eta_{\alpha \beta}$,
the matching condition reduces exactly to that of~\cite{JohnsonMcDaniel:2009dq}.
Taking into account the position of BH1, we then have
\begin{eqnarray}
(X^{\alpha})_{0} = x^{\alpha} - \frac{m_{2}}{m} b \,\hat x^{\alpha} = \tilde x^{\alpha} \,.
\label{eq:0thCT}
\end{eqnarray}

\subsubsection{First-Order: $O[(m_{2}/b)^{1/2}]$}

At first order, we have
\begin{eqnarray}
(g^{\NZ}_{\alpha \beta})_{1} =&
  (A_{\alpha}{}^{\gamma})_{0} (A_{\beta}{}^{\delta})_{0} (g^{\IZ}_{\gamma \delta})_{1}
+ 2\,(A_{(\alpha}{}^{\gamma})_{1} (A_{\beta)}{}^{\delta})_{0} (g^{\IZ}_{\gamma \delta})_{0}
\,,
\end{eqnarray}
and using $(g^{\NZ}_{\alpha \beta})_{1} = (g^{\IZ}_{\alpha \beta})_{1} = 0$,
$(A_{\alpha}{}^{\beta})_{0} = \delta_{\alpha}{}^{\beta}$,
and $(g^{\IZ}_{\gamma \delta})_{0} = \eta_{\alpha \beta}$, the above equation becomes
\begin{eqnarray}
(A_{(\alpha \beta)})_{1} = 0 \,.
\end{eqnarray}
Equation~(5.6) of~\cite{JohnsonMcDaniel:2009dq} can be used
to find the general expression of $(X_{\alpha})_{1}$:
\begin{eqnarray}
(X_{\alpha})_{1} = (F_{\alpha \beta})_{1} x^{\beta} + (D_{\alpha})_{1}
= (F_{\alpha \beta})_{1} \tilde x^{\beta} + (C_{\alpha})_{1} \,,
\label{eq:1stCT}
\end{eqnarray}
where $(F_{\alpha \beta})_{1}$ is a constant $4 \times 4$ antisymetric matrix,
$(C_{\alpha})_{1}$ (also $(D_{\alpha})_{1}$) is a constant $4 \times 1$ matrix
and we have used Eq.~\eqref{eq:0thCT}.
This is the most general solution of the flat-space Killing equation.

\subsubsection{Second-Order Matching: $O[(m_{2}/b)^1]$}

At second order, using $(g^{\IZ}_{\alpha \beta})_{1} = 0$, we have
\begin{eqnarray}
\fl
(g^{\NZ}_{\alpha \beta})_{2} =&
  (A_{\alpha}{}^{\gamma})_{0} (A_{\beta}{}^{\delta})_{0} (g^{\IZ}_{\gamma \delta})_{2}
+ (A_{\alpha}{}^{\gamma})_{1} (A_{\beta}{}^{\delta})_{1} (g^{\IZ}_{\gamma \delta})_{0}
+ 2(A_{(\alpha}{}^{\gamma})_{2} (A_{\beta)}{}^{\delta})_{0} (g^{\IZ}_{\gamma \delta})_{0} \,,
\end{eqnarray}
which gives us
\begin{eqnarray}
(g^{\NZ}_{\alpha \beta})_{2} &= (g^{\IZ}_{\alpha \beta})_{2}
+ (F_{\alpha}{}^{\gamma})_{1} (F_{\beta \gamma})_{1} + 2(A_{(\alpha \beta)})_{2} \,.
\label{eq:2ndM}
\end{eqnarray}
When we define
\begin{eqnarray}
S_{\alpha \beta} &= A_{(\alpha \beta)} = \partial_{(\alpha} X_{\beta)} \,,
\end{eqnarray}
Eq.~\eqref{eq:2ndM} can be rewritten as
\begin{eqnarray}
\label{matching-eq}
2(S_{\alpha \beta})_{2} &= (g^{\NZ}_{\alpha \beta})_{2}
- (g^{\IZ}_{\alpha \beta})_{2} - (F_{\alpha}{}^{\gamma})_{1} (F_{\beta \gamma})_{1} \,,
\end{eqnarray}
which is the matching equation one must solve at second-order.

Based on~\cite{JohnsonMcDaniel:2009dq}, the above equation is integrable
if the Riemann tensor associated with $S_{\alpha \beta}$ is null at second order, i.e.,
\begin{eqnarray}
{\cal I}_{\alpha\beta\gamma\delta} = \partial_{\alpha \beta}  (S_{\gamma \delta})_{2}
+ \partial_{\gamma \delta} (S_{\alpha \beta})_{2}
- \partial_{\alpha \delta} (S_{\gamma \beta})_{2}
- \partial_{\gamma \beta} (S_{\alpha \delta})_{2}
= 0 \,,
\label{eq:integ}
\end{eqnarray}
for all sets of $\alpha$, $\beta$, $\gamma$ and $\delta$. When we take $\gamma = \delta = 0$
and $\alpha, \beta = i,j \not = 0$, the integrability condition becomes
\begin{eqnarray}
\partial_{ij}  (S_{00})_{2} = 0 \,.
\end{eqnarray}
By linear independence, we can split this equation into a polynomial and a non-polynomial part.
The latter (the one that diverges when $|\tilde{x}_{i}| \rightarrow 0$) gives
\begin{eqnarray}
\partial_{ij} \left(2\, \frac{(M)_{0}}{m_{2}} \frac{b}{(R)_{0}}\right)
- \partial_{ij} \left(2\, \frac{m_{1}}{m_{2}} \frac{b}{(r_{1})_{0}}\right) = 0 \,,
\end{eqnarray}
where we have introduced the notation $(r_{1}^{k})_{0} = \tilde{x}^{k}$ from the zeroth-order matching.
This equation then forces $(M)_{0} = m_{1}$. For the polynomial part, it is easy to verify
that constant and first order pieces (in the sense of a polynomial decomposition $a + bX + cX^{2} + ...$)
give consistent, although trivial, equations. The quadrupolar piece gives
\begin{eqnarray}
\partial_{ij} \Bigl[ 3 [(\mathbf{r}_{1})_{0} \cdot (\mathbf{\hat b})_{0} ]^{2} - [(\mathbf{r}_{1})_{0}]^{2}
+ (\bar{\cal E}_{kl})_{0} (X^{k})_{0} (X^{l})_{0} \Bigr] = 0 \,,
\end{eqnarray}
which then forces
\begin{eqnarray}
(\bar{\cal E}_{ij})_{0} &= \delta_{ij} - 3 \hat{x}_{i} \hat{x}_{j} \,.
\end{eqnarray}
This is consistent with Eq.~(5.17) of~\cite{JohnsonMcDaniel:2009dq}.
We have verified that different choices of $\alpha$, $\beta$, $\gamma$ and $\delta$
are consistent (see~\ref{app:fullV}).

Once we have determined that the matching equation is integrable, we can proceed to solve it
to find the coordinate transformation. Solving Eq.~\eqref{matching-eq} we find
\begin{eqnarray}
\fl
(X_{\alpha})_{2} =&
\Bigl[1-\frac{\tilde{x}}{b} \Bigr] \Delta_{\alpha \beta} \tilde{x}^{\beta}
+ \frac{\Delta_{\gamma \delta} \tilde{x}^{\gamma} \tilde{x}^{\delta}}{2b} \hat{x}_{\alpha}
+ \frac{1}{3b^{2}} \Bigl[(\tilde{x}^{k} \tilde{x}_{k})^{3/2}
- 3  \tilde{x}^{2} \sqrt{\tilde{x}^{k} \tilde{x}_{k}} \Bigr] \hat t_\alpha
\nonumber \\
\fl &
- \frac{1}{b^{2}} \Bigl[ (\tilde{x}^{k} \tilde{x}_{k}) \tilde{x} \hat{x}_{i}
- \tilde{x}^{2} \tilde{x}_{i} \Bigr] \delta_{\alpha}^{i}
- \frac{1}{2} (F_{\alpha}{}^{\gamma})_{1} (F_{\delta \gamma})_{1} \tilde{x}^{\delta}
+ (F_{\alpha \beta})_{2} \tilde x^{\beta} + (C_{\alpha})_{2} \,,
\label{eq:2ndCT}
\end{eqnarray}
as we show in detail in~\ref{app:fullV}.

The quantities $(F_{\alpha \beta})_{1}$, $(F_{\alpha \beta})_{2}$ and $(C_{\alpha})_{2}$
are determined when one carries out matching to higher order~\cite{JohnsonMcDaniel:2009dq}.
It should be noted, however, that since the IZ metric we employ here is in a different gauge
than that used in~\cite{JohnsonMcDaniel:2009dq},
the matching transformation is also different. Thus, one cannot simply use the results
of~\cite{JohnsonMcDaniel:2009dq} here.
For simplicity, we will set $(F_{\alpha \beta})_{2}=(C_{\alpha})_{2}=0$ in the next section,
while $(F_{\alpha \beta})_{1}$ is determined in~\sref{subsec:ext}.

\subsection{Matching in the Spinning Case}

Let us now concentrate on matching in the spinning case and begin by estimating the order
at which the spinning contributions would enter the matching calculations.
Spin terms first enter the NZ metric at ${\cal{O}}(v^5)$, ${\cal{O}}(v^4)$ and ${\cal{O}}(v^5)$
in the $g_{00}$, $g_{0i}$ and $g_{ij}$ components, respectively. Spin terms first enter the IZ metric
at ${\cal{O}}(M A^2/R^3)={\cal{O}}(v^6)$, ${\cal{O}}(MA/R^2)$
and $O[M (\sqrt{M^2-A^2}-M)/R^2]={\cal{O}}(v^4)$,
and ${\cal{O}}(M A^2/R^3)={\cal{O}}(v^6)$ in the $g_{00}$, $g_{0i}$ and $g_{ij}$ components, respectively.
Here, since $A$ has dimensions of mass, $A/R$ is at most of ${\cal{O}}(M/R) = {\cal{O}}(v^{2})$.
Also, the leading-order part of the coupling between $z_{R/I,m}$ and
the BH spin arises at ${\cal{O}}(A R \,z_{R/I,m})={\cal{O}}(v^4)$,
since $z_{R/I,m} \propto m_{2}/b^{3}$.

This order counting argument suggests that spin contributions will first enter the matching calculation
at $O[(m_2/b)^{2}]={\cal{O}}(v^4)$. Therefore, one could take the spinning IZ metric
and the spinning NZ metric, and apply the non-spinning matching transformation to obtain a spacetime
that is properly asymptotically matched, without having to modify the matching transformations with spin terms.
This is one the main results of this paper.
We should note though that for many applications, the IZ metric must be written in appropriate harmonic,
and horizon-penetrating coordinates, while the perturbation should be constructed
with the appropriate boundary conditions~\cite{Yunes:2005ve}.

\subsection{Restoring Temporal Dependence of the Tidal Fields}\label{subsec:ext}

During matching, all fields are expanded in the BZ, and thus, quantities that are time-dependent,
such as the tidal fields, are Taylor expanded about $t = 0$ in $t/b \ll 1$.
However, one can restore the time-dependence of these tidal fields, as discussed
in Appendix B of~\cite{JohnsonMcDaniel:2009dq} via the replacements
\begin{eqnarray}
\hat{x}_{i} &\to \hat{x}_{i} \cos \omega t + \hat{y}_{i} \sin \omega t \,,
\quad
\hat{y}_{i} &\to - \hat{x}_{i} \sin \omega t + \hat{y}_{i} \cos \omega t \,.
\label{eq-subs}
\end{eqnarray}

Using the above substitution, the zeroth order coordinate transformation becomes
\begin{eqnarray}
(X_{\alpha})_{0} &= x_{\alpha} - \frac{m_{2}}{m} b \,(\hat{x}_{\alpha} \cos \omega t + \hat{y}_{\alpha} \sin \omega t)
\nonumber \\
&= x_{\alpha} - \frac{m_{2}}{m} b \,\hat{x}_{\alpha}
- \sqrt{\frac{m_2}{b}}\,\sqrt{\frac{m_2}{m}}\, t\, \hat{y}_{\alpha} + {\cal{O}}(v^2) \,.
\end{eqnarray}
One can think of $\hat{x}_{\alpha}$ as a zeroth-order term, while $\hat{y}_{\alpha}$ is a first-order term.
It is noted that the latter is of the same form as Eq.~\eqref{eq:1stCT} if we rewrite it as
$- \sqrt{m_2/m}\, t\, \hat{y}_{\alpha} = \sqrt{m_2/m}\, \hat{t}_{\beta}\, \hat{y}_{\alpha} \tilde{x}^{\beta}$
where $\hat{t}_{0}=-1$. We then have
\begin{eqnarray}
(F_{\alpha \beta})_{1}
&= 2\, \sqrt{\frac{m_2}{m}}\, \hat{t}_{[\beta}\, \hat{y}_{\alpha]} \,,
\end{eqnarray}
where we used the antisymmetrization properties of the $(F_{\alpha \beta})_{1}$ matrix
that is necessary to leave the Minkowski metric in the zeroth-order matching
unchanged in this coordinate transformation. As noted in~\cite{JohnsonMcDaniel:2009dq},
$(F_{\alpha \beta})_{1}$ represents a boost
(see also the low-order matching result of~\cite{Reifenberger:2012yg}).
Therefore, we may derive the first-order time coordinate transformation from a Lorentz boost.

Using the substitutions in Eq.~\eqref{eq-subs}, the quadrupolar field becomes
\begin{eqnarray}
z_{R,0} = \frac{2 m_2}{b^3} \,,
\quad
z_{R,2} = \frac{6 m_2}{b^3} \cos 2\omega t \,,
\quad
z_{R,-2} = \frac{6 m_2}{b^3} \cos 2\omega t  \,,
\nonumber \\
z_{I,2} = - \frac{6 m_2}{b^3} \sin 2\omega t \,,
\quad
z_{I,-2} = \frac{6 m_2}{b^3} \sin 2\omega t \,,
\end{eqnarray}
in terms of $z_{R/I,m}$ and the other components vanish.

\section{Numerical Analysis}\label{sec:numerics}

Using the matched metrics of the previous sections,
and the transition functions discussed in~\ref{app:trans},
we can construct an approximate global metric. In this section, we study equal-mass
BBHs and focus on the IZ and NZ only. The FZ metric becomes non-negligible
at field points $r/m \gtrsim 40$ and $110$ for orbital separations of $b=10m$ and
$20m$ (see Eq.~\eqref{eq:transFZ}), and it is automatically asymptotically matched
to the NZ one by construction.

\begin{figure*}[htb]
\begin{center}
\includegraphics[width=0.8\textwidth,clip=true]{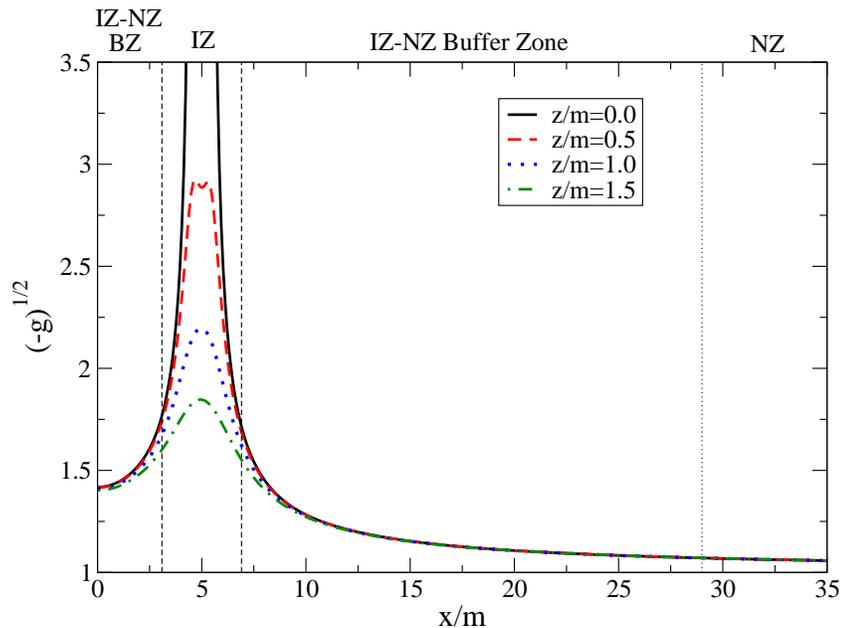}
\end{center}
\caption{\label{fig:vol_spin_mat}
Global $\sqrt{-g}$ for an equal-mass BBH with equal and aligned spins $(\chi_1,\chi_{2})=(0.9,0.9)$.
The orbital separation is $b=10m$, and each BH is located on the $x$-axis at $\pm 5 m$.
The volume element is plotted as a function of $x$, with $y=0$ and different colors and styles
correspond to different $z$-values. Since the figure is symmetric about $x = 0$, we only present results
around BH1. The vertical dashed and dotted lines roughly correspond to the inner and outer boundaries
of the IZ/NZ transition, respectively when $z/m=0$. In the region outside the vertical dotted lines,
the global metric reduces essentially to the NZ one, while in the region inside the dashed lines
around $x/m=5$ the global metric reduces to the IZ one.
Elsewhere, the global metric transitions between IZ and NZ metrics.
}
\end{figure*}

Let us first look at the volume element of the 4-metric for a BBH
with spins $\chi_1=0.9=\chi_2$, where the dimensionless spin parameters
$\chi_A \equiv |\vec{s}_A|/m_A^2$.
Although the volume element is coordinate dependent,
this is still a useful quantity to study when matching to verify that indeed
the metrics approach each other smoothly in the BZ and in the given coordinate system.
Figure~\ref{fig:vol_spin_mat} plots the volume element
as a function of $x$ for different values of $z$ and $y=0$. This figure shows
that indeed the IZ and NZ metrics smoothly match onto each other.
The smooth matching exhibited by the volume element is characteristic of
all metric components.

\begin{figure*}[htb]
\begin{center}
\begin{tabular}{l}
\includegraphics[width=0.5\textwidth,clip=true]{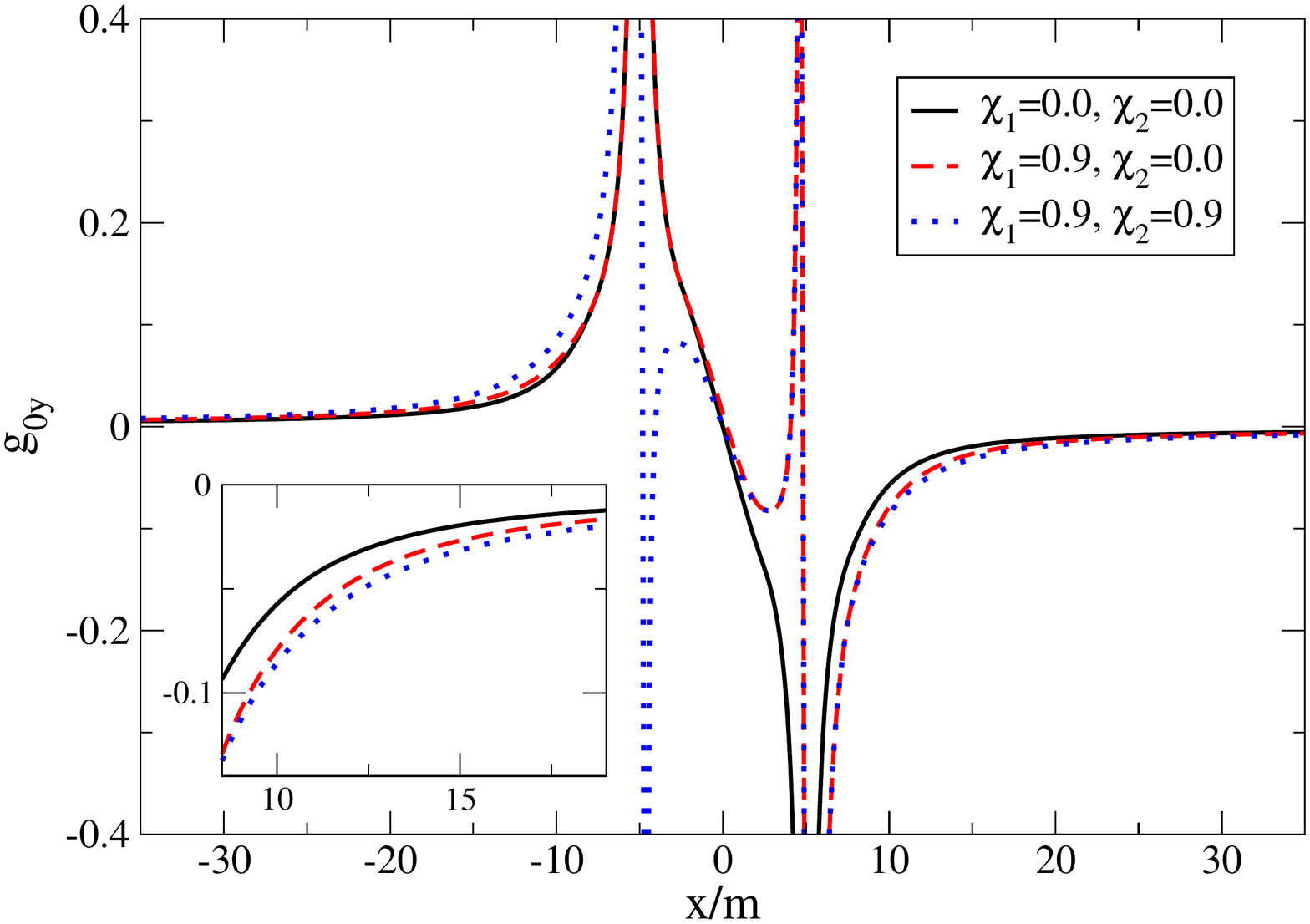}
\includegraphics[width=0.5\textwidth,clip=true]{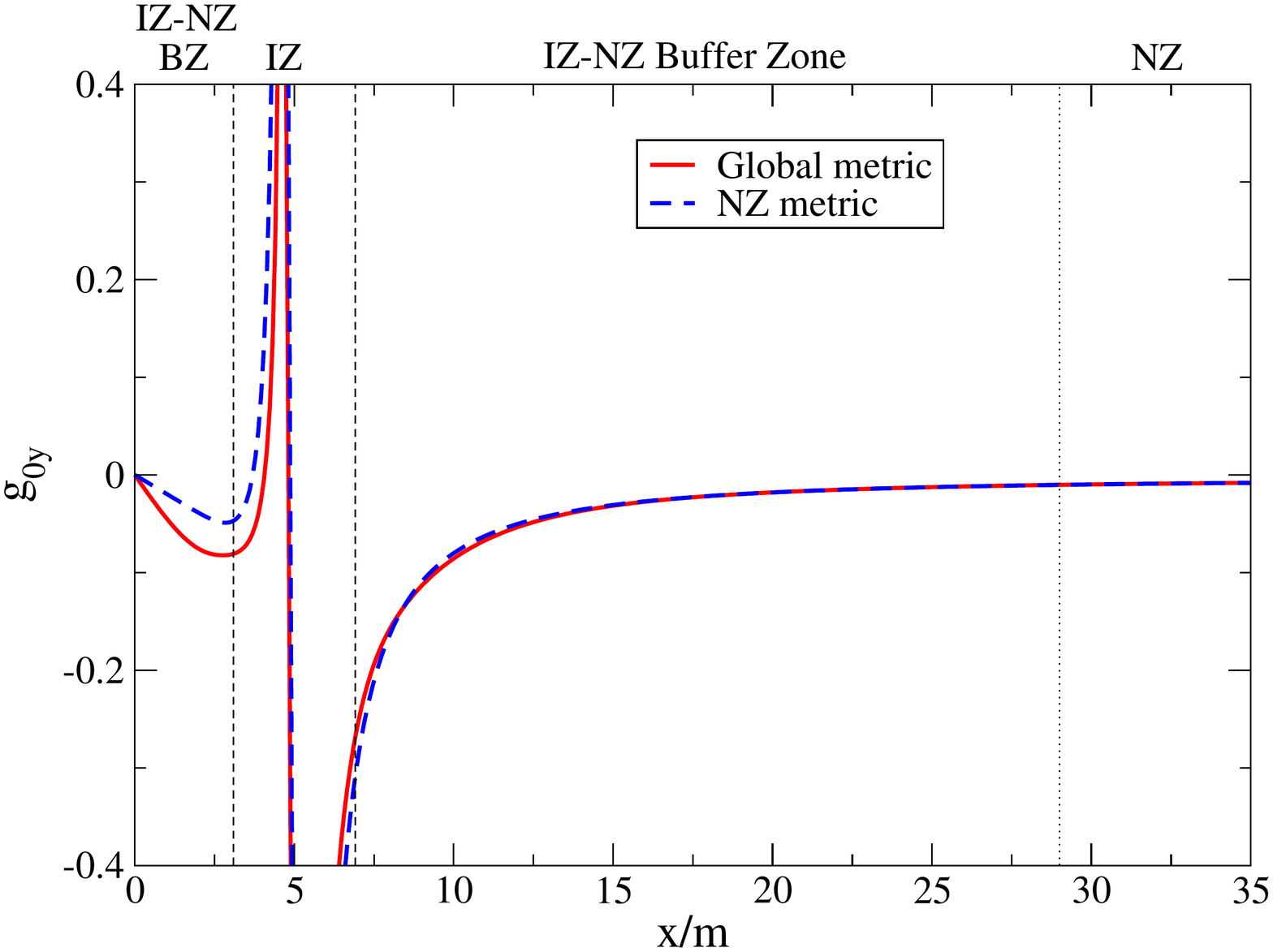}
\end{tabular}
\end{center}
\caption{\label{fig:gty_spin}
Left: Global $g_{0y}$ along the $x$-axis for an equal-mass BBH with different spins.
The inset zooms in to the middle region of the transition from the IZ to the NZ.
The IZ1 metric contributes at $\sim 2\%$ to the global metric $x/m=15$.
Right: Global and NZ only $g_{0y}$ along the $x$-axis for an equal-mass BBH with spins
$(\chi_1,\chi_{2}) = (0.9,0.9)$. The vertical lines are the same as in Figure~\ref{fig:vol_spin_mat}.
}
\end{figure*}

\begin{figure*}[htb]
\begin{center}
\begin{tabular}{l}
\includegraphics[width=0.5\textwidth]{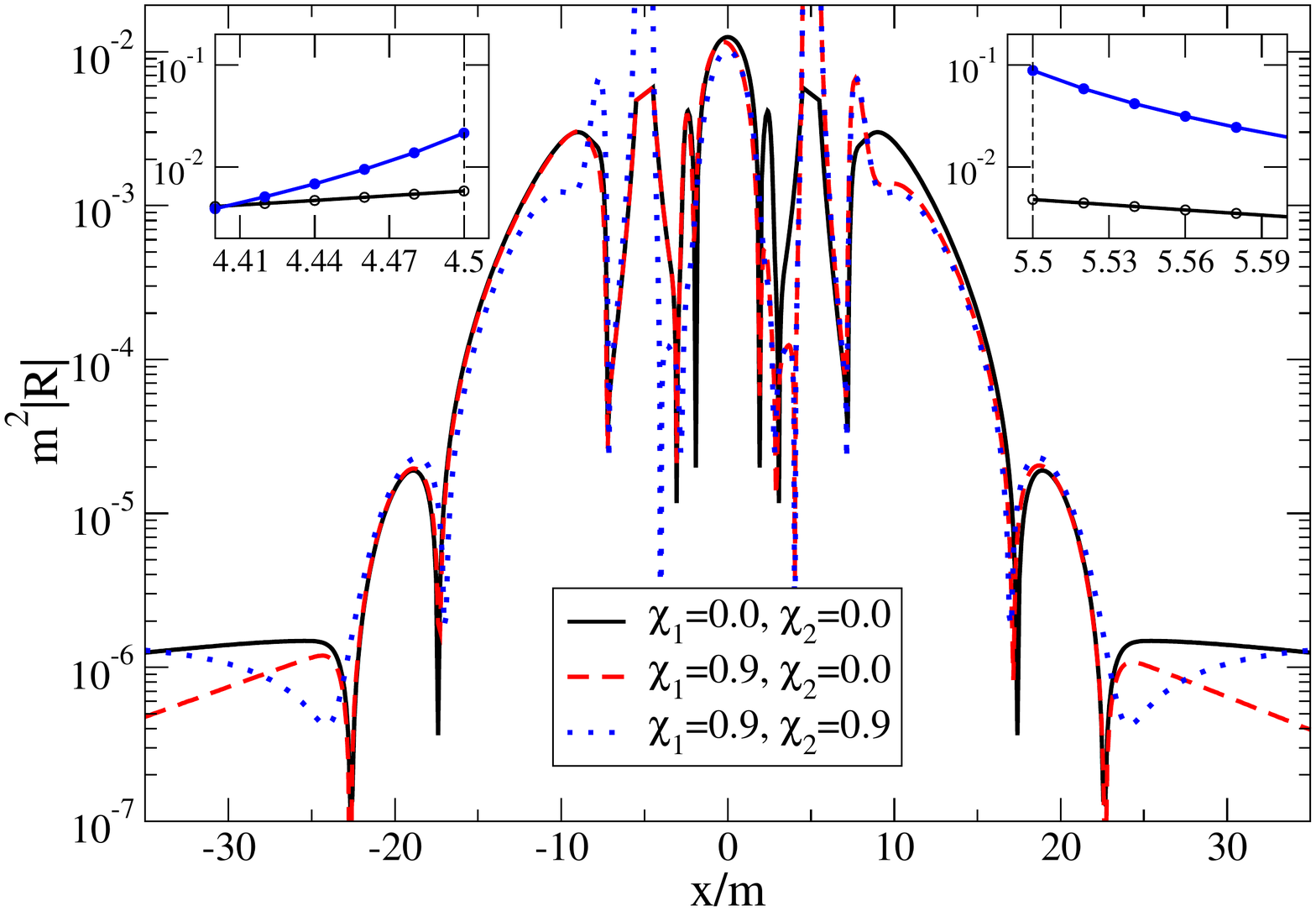}
\includegraphics[width=0.5\textwidth]{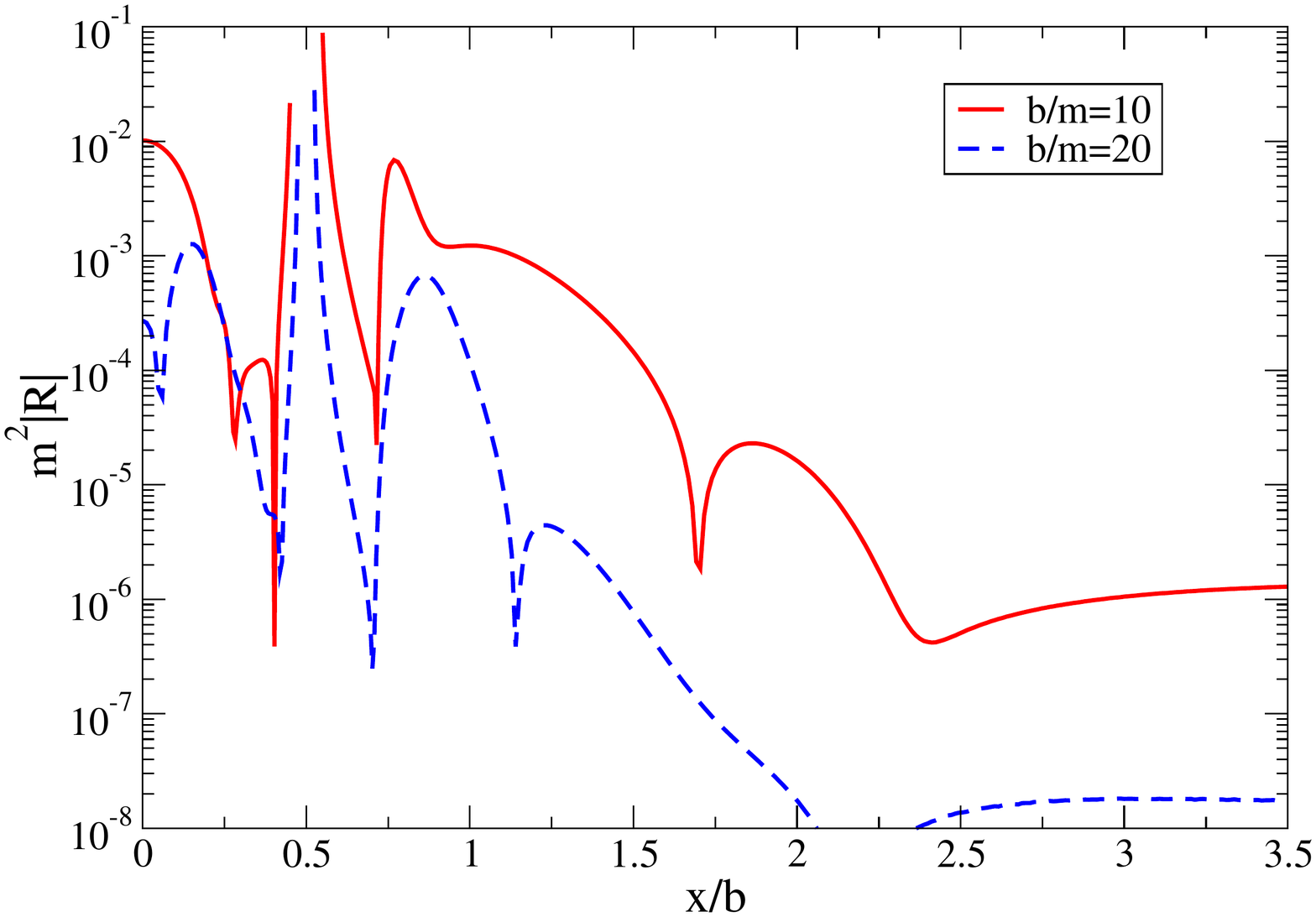}
\end{tabular}
\end{center}
\caption{\label{fig:Ricci_spin}
Left: Ricci scalar $R$ calculated with the global metric along the $x$-axis for an equal-mass BBH
with orbital separation $b = 10 m$ but different spins.
The insets zoom to a region close to BH1 for the $(\chi_1,\chi_{2}) = (0,0)$ and
$(\chi_1,\chi_2)=(0.9,0.9)$ cases. In the insets, the dashed vertical lines denote the location of
the event horizon of the IZ perturbed Kerr metric at $x=(5 \pm 0.5)m$, and we stop plotting the Ricci
scalar inside this region.
Right: Same as left panel but for the $(\chi_1,\chi_2)=(0.9,0.9)$ case plotted
for different orbital separations: $b=10m$ and $20m$.
The $x$-axis is the field point distance, normalized to the orbital separation.
}
\end{figure*}

Let us now study how the matching behaves as a function of spin parameter.
For this, we concentrate on the $g_{0y}$ component of the global metric, as this is one of the
components most affected by different spin values. The left panel of Figure~\ref{fig:gty_spin}
plots this component along the $x$-axis, for an equal-mass BBH
with different spin configurations: $(\chi_1,\chi_2)=(0,0)$, $(\chi_1,\chi_{2})=(0.9,0)$,
and $(\chi_1,\chi_{2})=(0.9,0.9)$. We observe the strong spin-dependence of this metric
component close to either BH (around $x/m = \pm 5$). In spite of this strong spin-dependence,
we observe that the matched metric smoothly transitions between the IZs and the NZ in the BZs.
Such a smooth transition is also shown on the right-panel of this figure, where we present
both the global and the NZ only metrics for the $(\chi_{1},\chi_{2}) = (0.9,0.9)$ case.

Finally, let us use the Ricci scalar $R$ as a measure of the accuracy to which the global metric
satisfies the vacuum Einstein equations. An exact solution would of course satisfy $R = 0$,
but since the metrics we are employing are approximate, their associated Ricci
scalars will not vanish exactly. The questions one wish to address are the following:
(i) are the IZ and NZ metrics as accurate as
they are supposed to be? (ii) do the transition functions introduce an error larger
than that intrinsically contained in the approximate IZ and NZ metrics? Question (i)
should clearly be answered in the affirmative,
if the IZ and NZ metrics have been correctly calculated. Question (ii)
should also be answered in the affirmative, provided the
transition functions are built to satisfy the Frankenstein theorems~\cite{Yunes:2006mx}.

The calculation of the Ricci scalar is not trivial due to the complexity of the IZ metric,
the matching parameter and coordinate transformation,
and the use of transition functions. For these reasons, we compute the Ricci on the $t=0$
spatial hypersurface through numerical methods.
In doing so, we not only need the spatial derivative of the global metric,
but also its time derivative. The latter requires knowledge of the time
evolution of the orbital phase and frequency, summarized in the Appendix of~\cite{Brown:2007jx},
as well as the location of each BH in Eq.~\eqref{eq:relCOM}.
We are using the currently-known highest PN results for the time evolution, i.e.,
the 3.5PN order for the non-spinning cases, and some lower order terms
for the spinning cases~\cite{Ajith:2012tt}.
When computing derivatives, it is of course critical to use sufficient numerical accuracy,
as otherwise the Ricci scalar will be highly inaccurate, specially in the region close to the BHs.

The left panel of Figure~\ref{fig:Ricci_spin} presents the Ricci scalar for different spin values,
while the right panel shows this scalar for different orbital separations. First, we see
that the metrics are indeed approximate solutions to the Einstein equations, as evidenced
by the right panel. That is, as the orbital separation is increased, the accuracy of the metrics
also increases and the Ricci scalar decreases at the expected rate.
Second, we see that the transition functions have been built to satisfy
the Frankenstein theorems, as they clearly do not introduce errors larger than those inherently
contained in the approximate metrics. That is, the height of the humps shown in the left panel of
Figure~\ref{fig:Ricci_spin} for $x \in [-15,-6] \cup [-4,4] \cup [6,15]$ is of the same
order as the error naturally inherent in the approximations. All of these conclusions are
mildly sensitive to the spin values, demonstrating the validity of the global metric for all spins.

The extreme Kerr limit, i.e., as $\chi_{A} \to 1$, must be treated carefully. From the matching
stand-point, this limit is perfectly well-behaved in a mathematical analysis sense. From a
numerical standpoint, however, this limit is difficult because of excision. When evolving
quantities numerically, one sometimes excises the region interior to the BH horizons,
placing the excision boundary somewhere well inside the horizon. In the extreme Kerr case,
the horizon shrinks and this could potentially lead to numerical problems with the excision boundary.
We stress however that this is a numerical problem, and not a mathematical one with asymptotic matching.

\section{Discussion}

We constructed a global approximate metric that represents the binary inspiral of
spinning compact objects. We split the spacetime into different zones: the IZs
(close to either BH), the NZ (far from either BH but less than a GW wavelength from
the center of mass) and the FZ (farther than a GW wavelength from the center of mass).
In each of these zones, the spacetime is approximated through either BH perturbation
theory techniques in the IZs or PN techniques in the FZs. These approximate metrics are
then related to each other via asymptotic matching, which provides a coordinate and
parameter transformation that renders adjacent metrics asymptotic to each other
in their mutually overlapping regions of validity. Once the metrics have been
asymptotically matched, they can be stitched together via certain transition functions
to yield an approximate metric for non-spinning BBHs~\cite{Yunes:2005nn,Yunes:2006iw,JohnsonMcDaniel:2009dq}.
Here, we have extended this work to spinning BBHs where the IZ metrics are
modeled through perturbed Kerr spacetimes~\cite{Yunes:2005ve} and the NZ
and FZ metrics include spin-orbit terms from PN theory~\cite{Tagoshi:2000zg,Faye:2006gx,Will:2005sn}.
After matching, the IZ metric reduces exactly to the Kerr solution in the limit of infinite binary separation.

The approximate global metric constructed in this way is technically valid on an initial
$t = 0$, spatial hypersurface, but it can be extended to capture the entire temporal evolution of the binary.
This evolution will be described in a future paper~\cite{Bruno_prep}.
Of course, this time-dependent, approximate global metric is only valid provided the approximate metrics
in each zone remain properly asymptotically matched, which in turn holds true if and only if a BZ exists.
These overlapping regions of validity are dynamically squeezed out as the binary shrinks,
and we expect the global construction to fail for sufficiently small separations,
as will be studied in~\cite{Bruno_prep}.
When the BHs are close enough, numerical relativity is required to describe the BBH spacetime.

We are planning to extend this work in multiple ways in the near future. We have here carried out
asymptotic matching to lowest order, and thus, a clear extension would be to re-do this calculation
but to higher order. For this to be feasible, however, one would first have to construct a vacuum-perturbed
Kerr metric that includes the first time derivative of the electric and magnetic quadrupole tensors,
as well as the octupole tensors. Such a metric is not available at this time,
which is why we were forced to stop the matching at lowest order.

Another avenue of future work would be to explore how the dynamical approximate global metric
constructed here affects certain astrophysical phenomena, such as accretion disks around a BBH.
A similar study was carried out in~\cite{Noble:2012xz}, except that there non-spinning BHs were considered
in the NZ metric. We expect spin-orbit coupling to be important to properly account for
the total angular momentum budget of the system. For example,
the spin-orbit coupling can have a dramatic effect on the inspiral rate. When the BH spins
are aligned (or even partially aligned) with the orbital angular momentum,
the merger is delayed, while when they are anti-aligned, the merger happens much more quickly.
This effect, also know  as ``hangup'' effect~\cite{Campanelli:2006uy},
is also responsible for very large kicks of the final merger remnant~\cite{Lousto:2011kp}.
We also expect spins to be important to properly describe the dynamics of the individual accretion
disks that may build-up around each BH during the inspiral process, specially at large separations.
It would be interesting to examine what the differences are when spin is included,
and whether this would lead to an electromagnetic observable precursor to BBH mergers.
We will explore this in a forthcoming paper.

\ack

We would like to thank Luc Blanchet, Nathan K. Johnson-McDaniel and Benjamin J. Owen
for useful comments and suggestions. L.G., H.N. and M.C. acknowledge support from NSF grants
AST-1028087 and PHY-0969855. L.G. thanks the Center for Computational Relativity and
Gravitation for supporting his Master 1 Research Internship.
N.Y. acknowledges support from NSF grant PHY-1114374, as well as support provided by the
National Aeronautics and Space Administration from grant NNX11AI49G,
under sub-award 00001944.
Some calculations used the computer algebra-systems MAPLE,
in combination with the GRTENSORII package~\cite{grtensor}.

\appendix

\section{Weyl Scalar for the IZ and NZ Metric}\label{app:comp}

Our starting point is the IZ metric, which is given by Eqs.~(YG-45) and~(YG-46),
where Eq.~(YG-n) denotes Eq.~(n) in~\cite{Yunes:2005ve}. Since the $m=0$ mode
is of higher-order in the analysis of~\cite{Yunes:2005ve}, it was ignored there.
In this paper, however, we wish to include this mode, and thus,
we derive it following the methods in~\cite{Yunes:2005ve}.

First, the radial function $R_m$ in Eq.~(YG-8) is rather simple when $m=0$:
\begin{eqnarray}
R_0 &= \mbox{const.}
= 1 \,,
\end{eqnarray}
where this constant is determined by requiring that $R_m \to 1$ for large radius.
One can also think of this solution as that for the $m \neq 0$ modes
in the Schwarzschild limit~\cite{Poisson:2004cw}. Using Eq.~(YG-21) for $R_0$,
we calculate the potential $\Psi$.
The radial dependence is simply $\Delta^2$, and we can directly extend Eq.~(YG-34)
to including the $m=0$ mode.
Thus, although there is a singular coefficient in the Schwarzschild limit (see Eq.~(YG-33)),
we may use Eq.~(YG-34) for the non-spinning
case\footnote{In the slow-motion approximation, i.e., when the characteristic
velocity $v \ll 1$, the time dependence of $z_{\ell m}$ in the potential $\Psi$ is also slow,
and we do not need to consider a rapidly spinning BH~\cite{Yunes:2005ve}.}.
If we considered time-dependent perturbations,
the reconstruction of the metric would be more complicated, as discussed
for example, in~\cite{Lousto:2002em,Ori:2002uv}.

Next, using Eq.~(YG-45) with all $(\ell,m)=(2,m)$ modes,
we calculate the Weyl scalar $\psi_0$.
One finds that $\psi_0 = -(1/2) \psi_0^{\rm (orig)}$ where $\psi_0^{\rm (orig)}$
denotes the Weyl scalar in Eq.~(YG-5). Therefore,
we have obtained a relation between ${\cal E}_{kl}$, ${\cal B}_{kl}$ and $z_{R/I,m}$,
as given in Eq.~\eqref{eq:zRItoEB}. The factor of $2$ difference arises due to a difference
in normalization of $\psi_0$ and $\Psi$ (see, e.g., (15) in~\cite{Whiting:2005hr}).
In practice, the time-time and time-space components shown in~\cite{Yunes:2005ve}
and \cite{JohnsonMcDaniel:2009dq} are the same in the $M/R \to 0$ limit with Eq.~\eqref{eq:zRItoEB}.

The Weyl scalar $\psi_0$ can be calculated directly from the NZ metric,
independently from \sref{sec:match}. Calculating the Weyl tensor from the NZ metric
and deriving the Weyl scalar by contracting with the tetrad,
we can compare it with the IZ $\psi_0$ in the BZ. From Eq.~\eqref{eq:NZ_bz}
for the NZ metric in the BZ, we have
\begin{eqnarray}
\psi_0 =&
\frac{3\,m_2}{2\,b^3}
[ ( 1+ \cos^2 \theta) \cos 2 \phi   - \sin^{2} \theta
-2\,i\cos \theta \sin 2 \phi ] \,,
\end{eqnarray}
to leading order in $m_2/b$, where we are using spherical polar coordinates
and we ignored any coordinate difference between the IZ and NZ.
Matching then forces $z_{R,0}=2m_2/b^3$, $z_{R,\pm2}=6m_2/b^3$
and all other coefficients to vanish. The above result is converted to
${\cal E}_{ij} = (m_2/b^3)(\delta_{ij} - 3 \hat{x}_{i} \hat{x}_{j})$ via Eq.~\eqref{eq:zRItoEB}.

\section{Detailed Verification of the Matching Calculation}\label{app:fullV}

We have so far studied matching for a given set of indices
in the integrability condition of Eq.~\eqref{eq:integ},
but now we extend this result to all indices.
Using the same notations as before, and taking
$(\alpha, \beta, \gamma) = (i,j,k) \neq 0$ and $\delta = 0$, we now have
\begin{eqnarray}
\partial_{i j}  (S_{k 0})_{2} - \partial_{k j} (S_{i 0})_{2} &= 0 \,,
\end{eqnarray}
which is explicitly written as
\begin{eqnarray}
\fl
0 =&
- \delta_{jk} \frac{\tilde{x}_{i}}{(R)_{0}^{3}} (\bar{\cal E}_{nm})_{0} (X^{n})_{0} (X^{m})_{0}
+ 2\delta_{jk} \frac{1}{(R)_{0}} (\bar{\cal E}_{in})_{0} (X^{n})_{0}
- 2  \frac{\tilde{x}_{j} \tilde{x}_{k}}{(R)_{0}^{3}} (\bar{\cal E}_{in})_{0} (X^{n})_{0}
\nonumber \\
\fl &
- \delta_{ij} \frac{\tilde{x}_{k}}{(R)_{0}^{3}} (\bar{\cal E}_{nm})_{0} (X^{n})_{0} (X^{m})_{0}
+ 3 \frac{\tilde{x}_{i} \tilde{x}_{j} \tilde {x}_{k}}{(R)_{0}^{5}} (\bar{\cal E}_{nm})_{0} (X^{n})_{0} (X^{m})_{0}
\nonumber \\
\fl &
- \delta_{ik} \frac{\tilde{x}_{j}}{(R)_{0}^{3}} (\bar{\cal E}_{nm})_{0} (X^{n})_{0} (X^{m})_{0}
+ 2 \delta_{ik} \frac{1}{(R)_{0}} (\bar{\cal E}_{jn})_{0} (X^{n})_{0}
- 2 \frac{\tilde{x}_{i} \tilde{x}_{k}}{(R)_{0}^{3}} (\bar{\cal E}_{jn})_{0} (X^{n})_{0}
\nonumber \\
\fl &
+ 2 \frac{\tilde{x}_{k}}{(R)_{0}} (\bar{\cal E}_{ij})_{0}
+ 2\delta_{ij} \frac{1}{(R)_{0}} (\bar{\cal E}_{kn})_{0} (X^{n})_{0}
+ 2 \frac{\tilde{x}_{j}}{(R)_{0}} (\bar{\cal E}_{ki})_{0}
+ 2 \frac{\tilde{x}_{i}}{(R)_{0}} (\bar{\cal E}_{kj})_{0}
\nonumber \\
\fl &
- 2 \frac{\tilde{x}_{i} \tilde{x}_{j}}{(R)_{0}^{3}} (\bar{\cal E}_{kn})_{0} (X^{n})_{0}
- (i \leftrightarrow k) \,,
\end{eqnarray}
where $\tilde{x}^{i}=(X^{i})_{0}$.
This equation is trivially consistent, and thus,
it provides no additional information about $(\bar{\cal E}_{ij})_{0}$.

The last set of indices to verify is $(\alpha, \beta, \gamma, \delta) = (i,j,k,l) \neq 0$.
The integrability condition becomes
\begin{eqnarray}
\partial_{ij}  (S_{kl})_{2} + \partial_{kl} (S_{ij})_{2}
- \partial_{il} (S_{kj})_{2}  - \partial_{kj} (S_{il})_{2} &= 0 \,.
\end{eqnarray}
As before, this equation can be divided into a polynomial and a non-polynomial part.
The non polynomial part gives again $(M_{1})_{0} = m_{1}$, and once more the constant
and first order pieces give trivial relations. The quadrupolar part gives
\begin{eqnarray}
\fl
0 =&
  (6 \hat{x}_{i} \hat{x}_{j} - 2 \delta_{ij})\delta_{kl}
+ \frac{2}{3} (\bar{\cal E}_{ij})_{0} \delta_{kl}
+ \frac{4}{3} (\bar{\cal E}_{kl})_{0} \delta_{ij}
+ (6 \hat{x}_{k} \hat{x}_{l} - 2 \delta_{kl})\delta_{ij}
+ \frac{2}{3} (\bar{\cal E}_{kl})_{0} \delta_{ij}
\nonumber \\
\fl &
+ \frac{4}{3} (\bar{\cal E}_{ij})_{0} \delta_{kl}
- (6 \hat{x}_{i} \hat{x}_{l} - 2 \delta_{il})\delta_{kj}
+ \frac{2}{3} (\bar{\cal E}_{il})_{0} \delta_{kj}
+ \frac{4}{3} (\bar{\cal E}_{kj})_{0} \delta_{il}
+ \frac{4}{3} (\bar{\cal E}_{il})_{0} \delta_{kj}
\nonumber \\
\fl &
- (6 \hat{x}_{k} \hat{x}_{j} - 2 \delta_{kj})\delta_{il}
+ \frac{2}{3} (\bar{\cal E}_{kj})_{0} \delta_{il} \,,
\end{eqnarray}
which once again leads to
$(\bar{\cal E}_{ij})_{0} = \delta_{ij} - 3 \hat{x}_{i} \hat{x}_{j}$.

Let us now focus on the coordinate transformation.
We need to solve
\begin{eqnarray}
2(A_{(\alpha \beta)})_{2} &= (g^{\NZ}_{\alpha \beta})_{2} - (g^{\IZ}_{\alpha \beta})_{2}
- (F_{\alpha}{}^{\gamma})_{1} (F_{\beta \gamma})_{1} \,,
\end{eqnarray}
or more explicitly, using the expression for $(\bar{\cal E}_{kj})_{0}$,
\begin{eqnarray}
\fl
2(A_{(\alpha \beta)})_{2} =&
\biggl[ (2-\frac{2}{b} \tilde{x})\Delta_{\alpha \beta} - (F_{\alpha}{}^{\gamma})_{1} (F_{\beta \gamma})_{1} \biggr]
- \biggl[ \delta_{\alpha}^{i} \delta_{\beta}^{j} \frac{2}{b^{2}}
\Bigl( (\tilde{x}^{k} \tilde{x}_{k}) \hat{x}_{i} \hat{x}_{j} - \tilde{x}^{2} \delta_{ij} \Bigr) \biggr]
\nonumber \\
\fl &
+ \biggl[ ( \delta_{\alpha}^{i} \hat t_\beta + \delta_{\beta}^{i} \hat t_\alpha ) \frac{1}{3b^{2}}
\Bigl( 3 \tilde{x}_{i} \sqrt{\tilde{x}^{k} \tilde{x}_{k}}
- 3 \frac{\tilde{x}_{i}}{(R)_{0}} \tilde{x}^{2} - 6 (R)_{0} \hat{x}_{i} \tilde{x} \Bigr) \biggr]
\,,
\label{eq:for2ndCT}
\end{eqnarray}
where we used the abbreviation $\tilde{x} = \tilde{x}^{k} \hat{x}_{k} = \tilde{x}^{\alpha} \hat{x}_{\alpha}$
because $\hat{x}_{0} = 0$.
The solution to this equation is obtained by adding the general flat solution to a particular solution.
The general solution is then
\begin{eqnarray}
(X_{\alpha})_{2,g} &= (F_{\alpha \beta})_{2} \tilde x^{\beta} + (C_{\alpha})_{2} \,.
\end{eqnarray}

The first bracket in Eq.~\eqref{eq:for2ndCT} is the same as the one found
in~\cite{JohnsonMcDaniel:2009dq} during second-order matching.
The particular solution for this term is
\begin{eqnarray}
(X_{\alpha})_{2,p1} =& \Bigl[1-\frac{\tilde{x}}{b} \Bigr] \Delta_{\alpha \beta} \tilde{x}^{\beta}
+ \frac{\Delta_{\gamma \delta} \tilde{x}^{\gamma} \tilde{x}^{\delta}}{2b} \hat{x}_{\alpha}
- \frac{1}{2} (F_{\alpha}{}^{\gamma})_{1} (F_{\delta \gamma})_{1} \tilde{x}^{\delta} \,.
\end{eqnarray}
For the second bracket, a particular solution is
\begin{eqnarray}
(X_{\alpha})_{2,p2} &= - \frac{1}{b^{2}} \Bigl[ (\tilde{x}^{k} \tilde{x}_{k}) \tilde{x} \hat{x}_{i}
 - \tilde{x}^{2} \tilde{x}_{i} \Bigr] \delta_{\alpha}^{i} \,.
\end{eqnarray}
For the third bracket, we can take the particular solution
\begin{eqnarray}
(X_{\alpha})_{2,p3} &= \frac{1}{3b^{2}} \Bigl[(\tilde{x}^{k} \tilde{x}_{k})^{3/2}
- 3  \tilde{x}^{2} \sqrt{\tilde{x}^{k} \tilde{x}_{k}} \Bigr] \hat t_\alpha \,.
\end{eqnarray}
Combining the general and particular solutions,
we obtain the expression in Eq.~\eqref{eq:2ndCT}.

\section{Transition function}\label{app:trans}

The construction of a smooth approximate global metric requires
the stitching of the asymptotically matched IZ, NZ and FZ metrics
through certain transition functions,
$f_{\rm far}$, $f_{\rm near}$, $f_{{\rm inner},1}$ and $f_{{\rm inner},2}$:
\begin{eqnarray}
g_{\mu\nu} =&
(1 - f_{\rm far})
\Bigl\{f_{\rm near} \bigl[f_{{\rm inner},1} \,g_{\mu\nu}^{\NZ}
+(1 - f_{{\rm inner},1} ) \,g_{\mu\nu}^{\rm (IZ1)}\bigr]
\nonumber \\ &
+ (1 - f_{\rm near} )\bigl[f_{{\rm inner},2} \,g_{\mu\nu}^{\NZ}
+(1 - f_{{\rm inner},2} ) \,g_{\mu\nu}^{\rm (IZ2)}\bigr]\Bigr\}
+ f_{\rm far} \,g_{\mu\nu}^{\FZ} \,.
\label{eq:wholemetric}
\end{eqnarray}
These transition functions can be modeled via
\begin{eqnarray}
\fl
f(r,\,r_0,\,w,\,q,\,s) =
\cases{
0, & \hspace{-85mm}$r \le r_0$, \\
\frac{1}{2} \left\{ 1 +
\tanh \left[\frac{s}{\pi} \left(\chi(r,\,r_0,\,w) - \frac{q^2}{\chi(r,\,r_0,\,w)} \right) \right] \right\}, & \\
\hspace{9mm}r_0 < r < r_0 + w, & \\
1, & \hspace{-85mm}$r \ge r_0 + w$,\\}
\end{eqnarray}
where $\chi(r, r_0, w)=\tan[\pi(r - r_0)/(2w)]$, and
$r_0$, $w$, $q$ and $s$ are parameters. Such a transition function has been greatly discussed
in~\cite{Yunes:2005nn,Yunes:2006iw,JohnsonMcDaniel:2009dq}
and it satisfies the Frankenstein conditions of~\cite{Yunes:2006mx}.

The transition functions are chosen with the following parameters.
\begin{eqnarray}
f_{\rm far} &= f(r,\,\lambda/5,\,\lambda,\,1,\,1.4) \,,
\\
f_{\rm near} &= f(x,\,2.2\,m_2-\frac{m_1 b}{m},\,b-2.2\,m,\,1,\,2.5) \,,
\\
f_{{\rm inner},A} &= f(r_A,\,0.4 \,r_A^T,\,3.5 \,(m b^4)^{1/5},\,0.2,\,b/m) \,.
\label{eq:transFZ}
\end{eqnarray}
Here, $r$ is the distance from the binary's center-of-mass to the field point.
The parameter $s=1.4$ has been discussed in~\cite{Bruno_prep}
and it is different from that of~\cite{JohnsonMcDaniel:2009dq},
which reduces the magnitude of the first radial derivative.
The quantity $\lambda = \pi\sqrt{b^3/M}$ is the gravitational wavelength
in the Newtonian limit. We assume that the BHs are initially located on the $x$-axis.

The transition function $f_{{\rm inner},A}$ is somewhat different from that
chosen in~\cite{JohnsonMcDaniel:2009dq}. This is because we match
here to a different order than in the latter paper. In particular,
the ``transition radius'' $r_A^T$, where the leading-order
uncontrolled remainder of adjacent zones becomes comparable, is here given by
\begin{eqnarray}
\left(\frac{m}{b}\right)\left(\frac{r_A^T}{b}\right)^3 &= \left(\frac{m_A}{r_A^T}\right)^2 \,,
\end{eqnarray}
where the left- and right-hand sides correspond to the IZ and NZ remainders,
respectively. The quantity $(r_A^T/b)^3$ in the IZ remainder arises due to ignorance
in the octupole contribution to the BH perturbation.
In the NZ, the remainder is $(m_A/r_A^T)^2$. The transition radius is thus
$r_A^T = (m_A^2 b^4/m)^{1/5}$, which is the same as in~\cite{Yunes:2006iw}.
The numerical coefficients in $f_{{\rm inner},A}$, $0.4$ and $3.5$, are
also from~\cite{Yunes:2006iw}, and the $(m b^4)^{1/5}$ dependence is obtained
by setting $m_A=m$ in $r_A^T$, as studied in~\cite{JohnsonMcDaniel:2009dq}.

\section{Higher Order Metrics}\label{app:higher}

Ref.~\cite{JohnsonMcDaniel:2009dq} constructed a 2.5PN, non-spinning NZ metric
following~\cite{Blanchet:1998vx}. In this paper, we have included certain higher PN order terms, 
when numerically calculating certain quantities in~\sref{sec:numerics}.

For the NZ metric, this higher-order extension is achieved by adding the following terms,
$\delta g_{\mu\nu}^{\NZ}$ to the 1.5PN order metric of Eq.~\eqref{eq:NZmetric}:
\begin{eqnarray}
\delta g_{00}^{\NZ} = \delta g_{00}^{\NZ,(6)} + \delta g_{00}^{\NZ,(7)} + {\cal{O}}(v^8) \,,
\nonumber \\
\delta g_{0i}^{\NZ} = \delta g_{0i}^{\NZ,(5)} + \delta g_{0i}^{\NZ,(6)} + {\cal{O}}(v^7) \,,
\nonumber \\
\delta g_{ij}^{\NZ} = \delta g_{ij}^{\NZ,(4)} + \delta g_{0i}^{\NZ,(5)} + {\cal{O}}(v^6) \,.
\end{eqnarray}
For example, $\delta g_{0i}^{\NZ,(5)}$ and $\delta g_{ij}^{\NZ,(4)}$ are explicitly given by
\begin{eqnarray}
\fl
\delta g_{0i}^{\NZ,(5)} =&
- n_{1}^{i} \Biggl[\frac{m_{1}^{2}}{r_{1}^{2}} (\mathbf{n}_{1} \cdot \mathbf{v}_{1})
+ \frac{m_{1} m_{2}}{\tilde{S}^{2}} \bigl\{16 (\mathbf{n}_{12} \cdot \mathbf{v}_{1}) - 12 (\mathbf{n}_{12} \cdot \mathbf{v}_{2})
+ 16 (\mathbf{n}_{2} \cdot \mathbf{v}_{1})
\nonumber \\
\fl
&
- 12 (\mathbf{n}_{2} \cdot \mathbf{v}_{2})\bigr\} \Biggr]
- n_{12}^{i} m_{1} m_{2} \Biggl[ 6 (\mathbf{n}_{12} \cdot \mathbf{v}_{12}) \frac{r_{1}}{b^{3}}
+ 4 (\mathbf{n}_{1} \cdot \mathbf{v}_{1}) \frac{1}{b^{2}} - 12 (\mathbf{n}_{1} \cdot \mathbf{v}_{1}) \frac{1}{\tilde{S}^{2}}
\nonumber \\
\fl
&
+ 16 (\mathbf{n}_{1} \cdot \mathbf{v}_{2}) \frac{1}{\tilde{S}^{2}}
- 4 (\mathbf{n}_{12} \cdot \mathbf{v}_{1}) \frac{1}{\tilde{S}} \Bigl(\frac{1}{\tilde{S}} + \frac{1}{b}\Bigr) \Biggr]
+ v_{1}^{i} \Biggl[ \frac{m_{1}}{r_{1}} (2 (\mathbf{n}_{1} \cdot \mathbf{v}_{1})^{2}
- 4 v_{1}^{2}) + \frac{m_{1}^{2}}{r_{1}^{2}}
\nonumber \\
\fl
&
+ m_{1} m_{2} \Bigl(\frac{3 r_{1}}{b^{3}} - \frac{2 r_{2}}{b^{3}}\Bigr)
- m_{1} m_{2} \Bigl( \frac{r_{2}^{2}}{r_{1} b^{3}} + \frac{3}{r_{1} b} - \frac{8}{r_{2} b} + \frac{4}{b \tilde{S}}\Bigr) \Biggr]
+ (1 \leftrightarrow 2)
\,,
\nonumber \\
\fl
\delta g_{ij}^{\NZ,(4)} =& 
\Biggl[ - \frac{m_{1}}{r_{1}} (\mathbf{n}_{1} \cdot \mathbf{v}_{1})^{2}
+ \frac{m_{1}^{2}}{r_{1}^{2}}
+ m_{1} m_{2} \Bigl(\frac{2}{r_{1} r_{2}} - \frac{r_{1}}{2 b^{3}}
+ \frac{r_{1}^{2}}{2 r_{2} b^{3}} - \frac{5}{2 r_{1} b}
+ \frac{4}{b \tilde{S}}\Bigr) \Biggr] \delta_{ij}
\nonumber \\
\fl
& 
+ 4 \frac{m_{1}}{r_{1}} v_{1}^{i} v_{1}^{j}
+ \frac{m_{1}^{2}}{r_{1}^{2}} n_{1}^{i} n_{1}^{j}
- 4 m_{1} m_{2} n_{12}^{i} n_{12}^{j} \left(\frac{1}{\tilde{S}^{2}} + \frac{1}{b \tilde{S}}\right)
\nonumber \\
\fl
&
+ 4 \frac{m_{1} m_{2}}{\tilde{S}^{2}} (n_{1}^{(i} n_{2}^{j)} + 2 n_{1}^{(i} n_{12}^{j)})
+ (1 \leftrightarrow 2)
\,.
\end{eqnarray}
In the above equations, we have introduced the notation
\begin{eqnarray}
\tilde{S} = r_{1} + r_{2} + b
\,,
\end{eqnarray}
where the quantity $\tilde{S}$ is a distance parameter that is not to be confused
with the magnitude of the spin angular momentum $s_{1}$ or $s_{2}$.
The other higher-order pieces can be obtained up from Eq.~(7.2) in~\cite{Blanchet:1998vx};
for example, $\delta g_{00}^{\NZ,(6)}$ is a contribution of ${\cal{O}}(1/c^6)$
that can be found in Eq.~(7.2a) of~\cite{Blanchet:1998vx}.

For the FZ metric, the higher-order extension is obtained by replacing
Eq.~\eqref{eq:FZ1} with
\begin{eqnarray}
g_{00}^{\FZ} = -\left[1 - \frac{1}{2}\, h^{00}_{\FZ} + \frac{3}{8}\, \left(h_{\FZ}^{00}\right)^{2} \right]
+ \frac{1}{2}\, h_{\FZ}^{kk} \,,
\nonumber \\
g_{0k}^{\FZ} = - \left[ 1 - \frac{1}{2}\, h_{\FZ}^{00} \right]h_{\FZ}^{0k} \,,
\nonumber \\
g_{kl}^{\FZ} = \left[1 + \frac{1}{2}\, h_{\FZ}^{00} - \frac{1}{8}\,\left(h_{\FZ}^{00}\right)^{2}
- \frac{1}{2} h_{\FZ}^{pp}\right]\delta^{kl} + h_{\FZ}^{kl} \,,
\end{eqnarray}
with ${\cal{O}}(v^6)$ remainders.
The metric potentials $h_{\FZ}^{\mu \nu}$ must also extended to higher-order via
\begin{eqnarray}
\fl
h_{\FZ}^{00} =
4\, \frac{{\cal{I}}}{r}
+ 2\, \partial_{kl} \left[ \frac{{\cal{I}}^{kl}(u)}{r} \right]
- \frac{2}{3}\, \partial_{klm} \left[ \frac{{\cal{I}}^{klm}(u)}{r} \right]
+ 7\, \frac{{\cal{I}}^{2}}{r^{2}} \,,
\nonumber \\
\fl
h_{\FZ}^{0k} =
- 2\, \partial_{l} \left[ \frac{\dot{\cal{I}}^{kl}(u)}{r} \right]
+ 2\, \epsilon^{lkp} \frac{n^{l} {\cal{J}}^{p}}{r^{2}}
+ \frac{2}{3}\, \partial_{lp} \left[ \frac{\dot{\cal{I}}^{klp}(u)}{r} \right]
+ \frac{4}{3}\, \epsilon^{lkp} \partial_{ls} \left[ \frac{{\cal{J}}^{ps}(u)}{r} \right] \,,
\nonumber \\
\fl
h_{\FZ}^{kl} = 2\,  \frac{\ddot{\cal{I}}^{kl}(u)}{r}
- \frac{2}{3}\, \partial_{p} \left[ \frac{\ddot{\cal{I}}^{klp}(u)}{r} \right]
- \frac{8}{3}\, \epsilon^{ps(k|} \partial_{s} \left[ \frac{\dot{\cal{J}}^{p|l)}(u)}{r} \right]
+ \frac{{\cal{I}}^{2}}{r^{2}} \hat{n}^k\hat{n}^l \,,
\end{eqnarray}
again with ${\cal{O}}(v^6)$ remainders.

\section*{References}

\bibliographystyle{iopart-num}
\bibliography{./references}

\end{document}